\documentclass[twocolumn,superscriptaddress,nobibnotes,aps,prd,showpacs,nofootinbib]{revtex4}
\usepackage{amssymb}
\usepackage{graphicx}

\def\be{\begin{equation}}
\def\ee{\end{equation}}
\def\bea{\begin{eqnarray}}
\def\eea{\end{eqnarray}}

\begin{document}

\title{Thin accretion disks around cold Bose-Einstein Condensate stars}
\author{Bogdan D\u{a}nil\u{a}}
\email{bogdan.danila22@gmail.com}
\affiliation{Department of Physics, Babes-Bolyai University, Kogalniceanu Street, Cluj-Napoca, Romania}
\author{Tiberiu Harko}
\email{t.harko@uccl.ac.uk}
\affiliation{Department of Mathematics, University College London, Gower Street,
London, WC1E 6BT, United Kingdom}
\author{Zolt\'{a}n Kov\'{a}cs}
\email{kovacsz2013@yahoo.com}
\affiliation{Max-Fiedler-Str. 7, 45128 Essen, Germany }
%\author{Gabriela Mocanu}
%\email{gabriela.mocanu@ubbcluj.ro}
%\affiliation{Astronomical Observatory, Cluj-Napoca, Romania}

\begin{abstract}
  Due to their superfluid properties some compact astrophysical objects, like neutron or quark stars, may contain a significant part of their matter in the form of a Bose-Einstein Condensate. Observationally distinguishing between neutron/quark stars and Bose-Einstein Condensate stars is a major challenge for this latter theoretical model. An observational possibility of indirectly distinguishing Bose-Einstein Condensate stars from neutron/quark stars is through the study of the thin accretion disks around compact general relativistic objects. In the present paper, we perform a detailed comparative study of the electromagnetic and thermodynamic properties of the thin accretion disks around rapidly rotating Bose-Einstein Condensate stars, neutron stars and quark stars, respectively. Due to the differences in the exterior geometry, the thermodynamic and electromagnetic properties of the disks (energy flux, temperature distribution, equilibrium radiation spectrum and efficiency of energy conversion) are different for these classes of compact objects. Hence in this preliminary study  we have pointed out some astrophysical signatures that may allow to observationally discriminate between Bose-Einstein Condensate stars and neutron/quark stars, respectively.

\end{abstract}


\date{\today }
\maketitle

%``

%\preprint{gr-qc/yymmnnn}

%%%%%%%%%%%%%%%%%%%%%%%%%%%%%%%%%%%%

\section{Introduction}\label{Sect.I}

Since its proposal by Bose \cite{Bose}, and generalization by Einstein \cite{Ein}, the quantum statistics of integer spin particles (bosons) did represent a fundamental field of study in both theoretical and experimental physics. One of the most important property of bosonic systems is their phase transition to a condensed state, in which all particles
are in the same quantum ground state. This quantum bosonic system is called a Bose-Einstein
Condensate (BEC), and from a physical point of view it is characterized by a sharp peak over a broader distribution
in both coordinate and momentum space. The quantum explanation for this behavior is that in a BEC particles become correlated with each
other, and their wavelengths overlap.  Note that the correlation means that the thermal wavelength $%
\lambda _{T}$ is greater than the mean inter-particle distance $l$. It occurs
at a critical temperature $T_c<2\pi \hbar ^{2}n^{2/3}/mk_{B}$, where
$m$ is the mass of an individual condensate particle, $n$ is the number
density, and $k_{B}$ is Boltzmann's constant \cite{Da99,
rev,Ch05,Pit,Pet,Zar}. A coherent quantum state in the bosonic system develops either when the particle density $\rho $
is high enough, or when the temperature $T$ is sufficiently low. From the experimental
point of view, the Bose-Einstein Condensation can be detected  by the observation of a sharp
peak in the velocity distribution, which always appears when the system is below the critical
temperature, $T<T_c$.

In the laboratory the Bose-Einstein Condensation was observed first in 1995 in
dilute alkali gases, such as vapors of rubidium and sodium. To obtain the condensation the gases were confined in a
magnetic trap, and cooled down to very low temperatures \cite{exp}. The experiment producing a laboratory BEC did represent a
major achievement in experimental condensed matter physics, and the
confirmation of the old and important predictions in theoretical statistical
physics of Bose and Einstein.  %.
%This shows that all the atoms have condensed into the same ground state, within a narrow peak in both momentum and coordinate space
\cite{exp}. Note that in recent years  quantum degenerate gases have been created by a multitude of experimental methods, including
combination of laser and evaporative cooling techniques. Thus the observation of the Bose-Einstein Condensation did open several new
lines of multidisciplinary research at the border of atomic, statistical and condensed matter
physics \cite{Da99,rev,Ch05,Pit,Pet,Zar,exp}.

The Bose-Einstein condensation processes are assumed to play an important role in the understanding of many fundamental processes in condensed matter physics. For example,  superfluidity of low temperature liquids, like $^{3}$He, can be explained by assuming a Bose-Einstein Condensation process \cite{Ch05}.
It is interesting to note that experimental observations as well as quantum theoretical calculations estimate the
condensate fraction $n_0$ at $T=0$ for superfluid helium to be
only around $n_{0}\approx 0.10$, a very small amount. Hence, since a strongly correlated pair of fermions
behaves approximately like a boson, the arising liquid helium superfluidity
%in $^{3}$He
can be interpreted as resulting from the Bose-Einstein Condensation of coupled fermions. A very similar physical model can be used to describe the
transition to a superconducting state in a solid material (metal). Superconductivity can thus be interpreted
as the condensation of electrons (or holes) into Cooper pairs. The formation of the Cooper pairs
drastically reduces the resistance caused by the motion of electrons in metals, and thus leads to the formation of a superconductor \cite{Ch05}.

Since Bose-Einstein condensation is a phenomenon that has been  observed and
thoroughly studied in terrestrial laboratories, the possibility that it may also occur in bosonic systems existing on
astrophysical or cosmological scales cannot be rejected \textit{a priori}. Thus, it was proposed
that dark matter, which is required to explain the dynamics of the neutral
hydrogen clouds at large distances from the centers of the galaxies, and which is assumed to be a
cold bosonic gravitationally bounded system, could also exist in the form of a Bose-Einstein Condensate \cite%
{Sin}.   A systematic study of
the properties of Bose-Einstein Condensed galactic dark matter halos was initiated in \cite%
{BoHa07}, and the astrophysical and cosmological implications of the existence of Bose-Einstein Condensed dark matter have been investigated in detail recently \cite{inv}.

From a theoretical point of view it was shown in \cite{BoHa07} that by introducing the Madelung representation of the wave function, the
dynamics of the Bose Einstein Condensate dark matter halo is described by the
continuity equation, and the hydrodynamic Euler equations of the standard classical fluid mechanics. Hence, Bose-Einstein Condensed
dark matter can be described theoretically as a gas, with the pressure and  density related by a barotropic equation of state. In the case of
self-interacting condensate dark matter with quartic self-interaction potential, the equation of state is polytropic
with index $n=1$ \cite{BoHa07}.

Bose-Einstein Condensation could play an important role in nuclear and quark matter physics, in the framework of the so-called
Bardeen-Cooper-Schrieffer (BCS) to Bose-Einstein Condensate (BEC) crossover.
 From a theoretical point of view it is expected that at ultra-high densities nuclear matter exists in the form of a
degenerate Fermi gas of quarks. The Cooper pairs of quarks form near the Fermi surface a Bose-Einstein
Condensate. Hence high density nuclear matter represents, from a physical point of view, a so called color superconductor \cite{csc}. When
the attractive interaction between fermions is strong enough, and the temperature drops below the critical temperature,
the fermions condense into the bosonic zero mode, and form a quark BEC
\cite{Bal95}. As a first step towards the formation of the BEC the fermions must form a BCS state, which can be realized  when the attractive
interaction between particles is weak. This system exhibits
superfluid properties, which are characterized by the existence of an energy gap for single particle
excitations. The energy gap is created by the formation of the Cooper pairs.
On the other hand a BEC is formed when the attractive interaction between fermions is
extremely strong. This interaction first leads to the formation of bound particles (bosons), which at some critical temperature $T_c$
start to condense into the bosonic zero mode. It is important to mention that the BCS and BEC states are smoothly
connected (crossover), without a phase transition between the two phases \citep{NiAb05}. For a recent review of the BCS-BEC crossover see \cite{revQ}.

The possibility of the existence of some forms of Bose-Einstein Condensates in neutron stars has been
considered a long time ago (see Glendenning \cite{Gl00} for a detailed discussion). One possibility for the formation of a BEC in a dense neutron star is the condensation of the negatively charged mesons, leading
to the  replacement of electrons with very high Fermi momenta by mesons \cite{Gl00}.
%The in-mediumvproperties of the $K^{-}$ mesons may be such that they could
%also condense in neutron matter.
Bose-Einstein Condensation of kaons and anti-kaons in compact objects
was also investigated in detail \cite{BaBa03,Ban04}. It turns out that the presence of pion or kaon condensates
may have at least two important effects on the global properties of dense neutron stars. Firstly, Bose-Einstein Condensation
softens the equation of state (EOS) of the stellar matter above the critical density for the onset of
condensation. An important consequence of the softening of the EOS is the reduction of the maximum neutron star mass. On the other hand, due to the
softening of the EOS, the central density increases significantly.  Secondly, the condensation of mesons would lead to considerably enhanced neutrino luminosities, much higher than those of normal neutron matter.
The increase in neutrino luminosity has important consequences on the neutron star cooling \citep{Gl00}. Another
particle, which may be present inside neutron stars, and which may form a Bose-Einstein Condensate, is the H-dibaryon. The H-dibaryon is a doubly strange six-quark composite. It has zero spin and isospin, and a baryon number $B=2$ \cite{Gl00}.
Neutron star matter may also contain an important fraction of $\Lambda $
hyperons, neutral subatomic hadrons. They  consist of one up, one down and
one strange quark, and they are labelled $\Lambda^{0}$ \cite{LambdaHyperon}. The $\Lambda $
hyperons may also combine to form H-dibaryons \cite{Hdibar}. Thus, H-matter
Bose-Einstein condensates may also exist at the center of very dense neutron stars, where the matter density is extremely high \citep{Gl00}.
An interesting possibility,  is that neutrino superfluidity, as suggested by Kapusta \cite{Ka04}, may also lead
to Bose-Einstein Condensation inside neutron stars \citep{Ab06}.

Thus theoretical results in nuclear matter physics indicate that the possibility of the existence of some forms of Bose-Einstein
Condensed matter inside compact astrophysical objects, like neutron or quark stars, or even the existence
of stars formed entirely from a pure BEC, cannot be excluded \textit{a priori}.
The properties of pure BEC stars have been considered in \cite{ChHa}. It was shown that stars formed of Bose-Einstein Condensates with particle masses of the order of two neutron masses (Cooper pair) and scattering length of the order of 10-20 fm have maximum masses of the order of $2 M_{\odot}$, maximum central densities of the order of $0.1-0.3\times 10^{16}$ g/cm$^3$ and minimum radii in the range of 10-20 km. Hence Bose-Einstein Condensed stars can form a large class of stable astrophysical objects, whose basic astrophysical parameters (mass and radius) sensitively depend on the mass of the condensed particle, and on the scattering length. It was also suggested that the  recently observed neutron stars with masses in the range of $2-2.4 M_{\odot}$ (Vela X-1, 4U 1700-377, and the black widow pulsar B1957+20) are Bose-Einstein Condensate stars. Further properties of BEC stars have been considered in \cite{Ch2}.

The structure of static and rotating BEC stars was investigated numerically in \cite{MT} by solving the Gross-Pitaevskii-Poisson system of coupled differential equations. It was shown,  with longer simulation runs, that within the computational limits of the simulation the BEC stars are stable. The physical properties of the self-gravitating Bose-Einstein Condensate were investigated in both non-rotating and rotating cases.

A fundamental theoretical problem is to find some clear astrophysical signatures, going beyond the global stellar parameters (mass and radius) that could indicate the presence of a BEC inside a neutron star, or to give a firm observational evidence for the existence of pure BEC stars. It is the purpose of the present paper to suggest that such a specific signature, indicating the presence of a pure BEC star, may indeed exist, and it can be obtained from the study of the radiation emission of thin accretion disks that usually form around compact general relativistic objects.

The growth of most astrophysical objects is determined by mass accretion. Due to the presence of interstellar matter,  accretion disks are generally formed around compact objects. Accretion disks are well known observationally, representing flattened
astronomical structures. They are  made of rapidly rotating hot gas, slowly spiraling onto a central massive and dense object. The gravitational energy of the gas motion is a source of heat, generated by the internal stresses and the dynamical friction. A small fraction of the heat is converted into radiation, which partially escapes. The radiation emission cools down the accretion disk. Therefore important information about the accretion disk physics comes from the radiation emitted from the disk. The radiation, detected in the radio, optical or X-ray frequency bands, allows astronomers to analyze
its electromagnetic spectrum, and its time variability.  Thus essential results about the physics of the disks can be obtained from observations.

 The cooling of the disk via the electromagnetic radiation emission from its surface represents an efficient mechanism that prevents the extreme heating of the disk. On the other hand, the thermodynamic equilibrium established in this way allows the disk to stabilize its thin vertical size. Usually the inner edge of the thin disk is located at the marginally stable orbit of the compact object potential. Hence in higher orbits the hot gas has a Keplerian motion \cite{PaTh74,Th74}. Since the electromagnetic radiation emission from the thin disk is determined by the external gravitational potentials, which in turn are determined by the EOS of the dense neutron or quark matter in the star, astrophysical observations of the emission spectra from accretion disks may lead to the possibility of
directly testing the equation of state of the dense matter inside compact general relativistic objects.

The emissivity properties of the accretion disks have been used to investigated, and obtain distinctive astrophysical signatures, for large classes of compact astrophysical objects, including naked singularities \cite{na}, gravastars \cite{gra} and wormholes \cite{wor}.  Specific electromagnetic disk signatures in different modified gravity theories, such as $f(R)$ gravity, brane world models, Chern-Simons models and the Horava-Lifshitz theory were considered \cite{mod}, while the properties of accretion disks around
 rotating and non-rotating neutron, quark, boson or fermion stars have been analyzed in \cite{Ko, To02,other}.

In the present paper we perform a comparative study of the rotational properties and of the disk emission properties for five neutron star EOSs, and of the quark matter bag model EOS with the polytropic $n=1$ BEC equation of state. The main goal of this preliminary investigation is to point out towards the possible existence of some observational signatures that may distinguish between these different classes of compact objects. We begin our analysis by considering the global rotational  properties of the considered compact objects. To obtain the equilibrium configurations of the rotating neutron, quark and BEC stars we use the
Rotating Neutron Star (RNS) code, as introduced in \cite{Ster}, and
discussed in detail in \cite{Ster1}. The RNS software provides
the metric potentials for different types of equations of state of compact rotating general
relativistic objects. We investigate three different cases, corresponding to stellar models with fixed mass and angular velocity, stellar models rotating at Keplerian frequencies, and stellar models with fixed central density and fixed polar radius to equatorial radius ratio.  As a next step in our study we use the exterior metrics  to obtain the physical
properties of the accretion disks for the considered equations of state.  Particular signatures appear in
the electromagnetic spectrum of the BEC stars, thus leading to the theoretical possibility of
directly testing, and discriminating,  the BEC equation of state of the dense matter by using
astrophysical observations of the emission spectra from accretion disks.

The present paper is organized as follows. In Section~\ref{sect1} we briefly review the basics of the Bose-Einstein Condensation and discuss the properties of Newtonian BEC stars. The electromagnetic and thermodynamic properties of accretion disks around compact general relativistic objects are described in Section~\ref{sect2}. The equations of state of the neutron, quark and BEC matter, as well as the global astrophysical properties of the considered stellar models are obtained and discussed in Section~\ref{sect3}. The electromagnetic and thermodynamic properties (flux, luminosity and temperature distribution) of the accretion disks around neutron, quark and BEC stars are obtained in Section~\ref{sect4}. We discuss our results and conclude our study in Section~\ref{sect5}.

\section{Bose Einstein Condensation}\label{sect1}

The most important characteristic of a quantum system of $N$ interacting bosons in the condensed state is that most of the particles
lie in the same single-particle quantum state. For a bosonic quantum system consisting of an extremely
large number of particles, the calculation of the physical parameters of the ground state
with the direct use of the Hamiltonian is generally very difficult, due to the high
computational cost. A significant simplification of the many-body type computations can be achieved with the use of some approximate methods.  One such approximate semi-analytic method is the mean field analysis of the quantum condensate. The basic idea of the mean field method is the
separation of the Bose-Einstein Condensate contribution to the total bosonic field operator. In the following in our analysis we assume that the bosonic astrophysical system consists of  scalar (zero spin) particles with non-zero
mass. From a simplifying physical point of view we assume that when the system makes a transition to a
Bose-Einstein Condensed phase, {\it the range of Van der Waals-type scalar mediated interactions among
particles becomes infinite}.

\subsection{The Gross-Pitaevskii equation}

In the second quantization approach the Hamiltonian describing a many body system of interacting bosons, confined by an
external potential $V_{ext}$, is given by
\begin{eqnarray}\label{ham}
&&\hat{H}=\int d\vec{r}\hat{\Phi}^{+}\left( \vec{r}\right) \left[ -\frac{\hbar
^{2}}{2m}\nabla ^{2}+V_{rot}\left( \vec{r}\right) +V_{ext}\left( \vec{r}%
\right) \right] \hat{\Phi}\left( \vec{r}\right) +\nonumber\\
&&\frac{1}{2}\int d\vec{r}d%
\vec{r}^{\prime }\hat{\Phi}^{+}\left( \vec{r}\right) \hat{\Phi}^{+}\left(
\vec{r}^{\prime }\right) V\left( \vec{r}-\vec{r}^{\prime }\right) \hat{\Phi}%
\left( \vec{r}\right) \hat{\Phi}\left( \vec{r}^{\prime }\right) ,
\end{eqnarray}
where $\hat{\Phi}\left( \vec{r}\right) $ and $\hat{\Phi}^{+}\left( \vec{r}%
\right) $ are the bosonic field operators, and $V\left( \vec{r}-\vec{r%
}^{\prime }\right) $ is the two-body interatomic potential, respectively \cite{Da99}-\cite{ Zar}.  The bosonic field operators annihilate and create a
particle at the position $\vec{r}$. $%
V_{rot}\left( \vec{r}\right) $ is the potential associated to the rotation
of the condensate.

In order to obtain  a significant
simplification of the mathematical formalism a number of approximate methods have been developed. One such approximate and simplifying semi-analytic approach is the mean field description of the condensate. In this approach
the condensate contribution to the bosonic field operator is separated out. For a uniform gas confined
in a volume $V$, BEC occurs in the single particle state $\Phi _{0}=1\sqrt{V}
$, having zero momentum. The field operator can then be decomposed  in the
form $\hat{\Phi}\left( \vec{r}\right) =\sqrt{N/V}+\hat{\Phi}^{\prime }\left(%
\vec{r}\right) $. By treating the operator $\hat{\Phi}^{\prime }\left( \vec{r%
}\right) $ as a small perturbation, the first order theory
for the excitations of the interacting Bose gases can be fully developed \cite{Da99}-\cite{Zar}.

In the general case of a non-uniform and time-dependent configuration the Heisenberg representation of the field operator  is given by
\be
\hat{\Phi}%
\left( \vec{r},t\right) =\psi \left( \vec{r},t\right) +\hat{\Phi}^{\prime
}\left( \vec{r},t\right) ,
\ee
where $\psi \left( \vec{r},t\right) $, also
called the condensate wave function, is the expectation value of the field
operator, $\psi \left( \vec{r},t\right) =\left\langle \hat{\Phi}\left( \vec{r%
},t\right) \right\rangle $. It is important to note that the wave function $\psi \left( \vec{r},t\right) $ is a classical field, and its absolute value
fixes the particle number density of the condensate through the relation $\rho \left( \vec{r}%
,t\right) =\left| \psi \left( \vec{r},t\right) \right| ^{2}$. On the other hand the
normalization condition for the condensate wave function is $N=\int \rho \left( \vec{r},t\right) d^{3}\vec{r}$%
, where $N$ is the total number of particles in the Bose-Einstein Condensate.

 The condensate wave function satisfies the following equation of motion, which can be obtained from
Heisenberg equation corresponding to the many-body Hamiltonian given by Eq.~(%
\ref{ham}),
\begin{eqnarray}
&&\hspace{-0.5cm}i\hbar \frac{\partial }{\partial t}\hat{\Phi}\left( \vec{r},t\right) =\left[
\hat{\Phi},\hat{H}\right] =
\left[ -\frac{\hbar ^{2}}{2m}\nabla
^{2}+V_{rot}\left( \vec{r}\right) +V_{ext}\left( \vec{r}\right)\right. +\nonumber\\
&&\left.\int d\vec{r%
}^{\prime }\hat{\Phi}^{+}\left( \vec{r}^{\prime },t\right) V\left( \vec{r}%
^{\prime }-\vec{r}\right) \hat{\Phi}\left( \vec{r}^{\prime },t\right) \right]
\hat{\Phi}\left( \vec{r},t\right).  \label{gp}
\end{eqnarray}

The zeroth-order approximation to the Heisenberg
equation is obtained  by replacing $\hat{\Phi}\left( \vec{r},t\right) $ with the condensate wave
function $\psi $. For short distances this is in general a
poor approximation for computing the integral containing the particle-particle interaction $%
V\left( \vec{r}^{\prime }-\vec{r}\right) $. However, it is important to note that in a dilute and cold gas,
at low energy only binary collisions are important. These collisions can be
characterized, independently of the exact form of the two-body potential, by a single physical  parameter, the $s$-wave scattering length $a$.
Therefore, one can obtain a very good approximation by
replacing the generally unknown potential $V\left( \vec{r}^{\prime }-\vec{r}\right) $ with an effective
interaction potential $V\left( \vec{r}^{\prime }-\vec{r}\right) =\lambda \delta \left(
\vec{r}^{\prime }-\vec{r}\right) $, with the coupling constant $\lambda $
determined by the scattering length $l_a$ via the relation $\lambda =4\pi \hbar ^{2}a/m$, where $m$ is the mass of the condensate particles.
With the use of the effective potential the integral in the bracket of Eq.~(%
\ref{gp}) can be easily calculated to give $\lambda \left| \psi \left( \vec{r},t\right) \right| ^{2}$.
Hence the resulting approximate equation of motion for the condensate is the Schr\"odinger equation with a quartic
nonlinear term \cite{Da99}-\cite{Zar}, also called the Gross-Pitaevskii equation. However, in order to obtain a more general and realistic
description of the Bose-Einstein Condensate stars, one may also  assume an
arbitrary non-linear self-interaction term of the form $g\left( \left| \psi \left( \vec{r},t\right)
\right| ^{2}\right)=g\left(\rho \right)$ \cite{Da99}.

Therefore the non-relativistic generalized Gross-Pitaevskii equation, describing a
gravitationally trapped rotating Bose-Einstein condensate, is given by
\begin{eqnarray}\label{sch}
i\hbar \frac{\partial }{\partial t}\psi \left( \vec{r},t\right) &=&\bigg[ -%
\frac{\hbar ^{2}}{2m}\nabla ^{2}+V_{rot}\left( \vec{r}\right) +V_{ext}\left(
\vec{r}\right) +\nonumber\\
&&g^{\prime }\left( \left| \psi \left( \vec{r},t\right)
\right| ^{2}\right) \bigg] \psi \left( \vec{r},t\right) ,
\end{eqnarray}
where we have denoted $g^{\prime }=dg/d\rho $.  As for $V_{ext}\left( \vec{r}%
\right) $, we assume that it is the gravitational potential, denoted for simplicity by $V$, $V_{ext}=V$.
The gravitational potential satisfies the Poisson equation of Newtonian gravity
\begin{equation}
\nabla ^{2}V=4\pi G\rho _{m},
\end{equation}
where $\rho _{m}=m\rho =m\left| \psi \left( \vec{r},t\right) \right| ^{2}$
is the mass density inside the Bose-Einstein condensate.

\subsection{The hydrodynamical representation}

The physical as well as the astrophysical properties of a Bose-Einstein Condensate, described by the
generalized Gross-Pitaevskii equation  Eq.~(\ref{sch}) can be
analyzed, computed and understood much more easily by using the so-called Madelung representation of the
wave function \cite{Da99}. In this representation we write the condensate wave function  $\psi $ in the specific form
\begin{equation}
\psi \left( \vec{r},t\right) =\sqrt{\rho \left( \vec{r},t\right) }\exp \left[
\frac{i}{\hbar }S\left( \vec{r},t\right) \right] ,
\end{equation}
where the function $S\left( \vec{r},t\right) $ has the physical dimensions of an
action. By substituting the above expression of  $\psi \left( \vec{r}%
,t\right) $ into Eq.~(\ref{sch}), it follows that the generalized Gross-Pitaevskii equation decouples into a system of two
first order partial differential equations for the two real functions $\rho _{m}$ and $\vec{v}$, respectively,
given by
\begin{equation}
\frac{\partial \rho _{m}}{\partial t}+\nabla \cdot \left( \rho _{m}\vec{v}%
\right) =0,  \label{cont}
\end{equation}
\begin{eqnarray}
&&\rho _{m}\left[ \frac{\partial \vec{v}}{\partial t}+\left( \vec{v}\cdot
\nabla \right) \vec{v}\right] =-\nabla P\left( \frac{\rho _{m}}{m}\right)-\nonumber\\
&&\rho _{m}\nabla \left( \frac{V_{rot}}{m}\right) -
\rho _{m}\nabla \left(
\frac{V_{ext}}{m}\right) -\nabla V_{Q},  \label{euler}
\end{eqnarray}
where we have introduced the quantum potential $V_Q$, defined as,
\begin{equation}
V_{Q}=-\frac{\hbar ^{2}}{2m}\frac{\nabla ^{2}\sqrt{\rho _{m}}}{\sqrt{\rho
_{m}}},
\end{equation}
the velocity of the Bose-Einstein Condensed fluid $\vec{v}$, given by
\begin{equation}
\vec{v}=\frac{\nabla S}{m},
\end{equation}
 and we have denoted
\begin{equation}
P\left( \frac{\rho _{m}}{m}\right) =g^{\prime }\left( \frac{\rho _{m}}{m}%
\right) \frac{\rho _{m}}{m}-g\left( \frac{\rho _{m}}{m}\right) .
\label{state}
\end{equation}

From its definition we can immediately see that the velocity field $\vec{v}$ is irrotational. Thus
it satisfies the condition $\nabla \times \vec{v}=0$. Therefore we obtain the important result that in the Madelung representation the equations
of motion of the Bose-Einstein Condensate in a gravitational field take the form
of the equation of continuity, and of the hydrodynamic Euler equations, respectively. Hence a
Bose-Einstein Condensate in an external gravitational field can be described as a gas whose
density and pressure are related by a barotropic equation of state %
\cite{Da99}-\cite{Zar}. The exact form of the equation of state of the BEC depends on the
non-linearity term $g$.

In the important case of a gravitationally bounded Bose-Einstein
Condensate with a very large number of particles the quantum pressure is important only near the external boundary of the quantum system. Hence in all astrophysically important situations the quantum pressure term
is much smaller than the non-linear interaction term \cite{BoHa07}.
%Thus the quantum
%stress term, corresponding to the quantum potential $V_Q$ can be neglected in the equation of motion of the condensate.
%This approximation is called the Thomas-Fermi approximation, and it has been extensively used for
%the study of the Bose-Einstein Condensates \cite{Da99}-\cite{Zar}. When the number of
%particles in the condensate becomes infinite, the Thomas-Fermi approximation
%becomes exact. The Thomas-Fermi approximation also corresponds to the
%classical limit of the theory, when one neglects all terms with
%powers of $\hbar $. Equivalently, it describes  the regime of strong repulsive interactions among
%particles. From a mathematical point of view, in the Thomas-Fermi approximation
%one neglects in the equation of motion all terms containing $%
%\nabla {\rho }$ and $\nabla {S}$.

In the most studied approach to the Bose-Einstein Condensates, the non-linearity
term $g$ is quadratic, and is given by
\begin{equation}
g\left( \rho _m\right) =\frac{u_{0}}{2}\left| \psi \right| ^{4}=\frac{u_{0}}{2}%
\rho _m^{2},
\end{equation}
where $u_{0}=4\pi \hbar ^{2}l_a/m$ \citep{Da99}-\cite{Zar}. The corresponding equation of
state of the condensate is
\begin{equation}\label{eqstate}
P\left( \rho \right) =K\rho ^{2}.
\end{equation}
%with
%\begin{equation}
%K=\frac{2\pi \hbar ^{2}a}{m^{3}} =0.1856\times 10^5 \left(\frac{a}{1\;{\rm fm}}\right)\left(\frac{m}{2m_n}\right)^{-3},
%\end{equation}
%where $m_n=1.6749\times 10^{-24}$ g is the mass of the neutron.
Therefore we have obtained the very important result that the equation of state of the standard Bose-Einstein Condensate with quartic
non-linearity is a polytrope, with index $n=1$.

In the following for simplicity we consider only the case of the Bose - Einstein condensates with quartic non-linearity. In this particular case the physical properties of the condensates are also relatively well known from numerous laboratory experiments. From theoretical point of view it is important to note that their properties can be described in terms of only two free parameters, the mass $m$ of the condensate particle, and the scattering length $a$, describing the particle interaction, respectively.

\subsection{Masses and radii of Newtonian static BEC stars}

Since the EOS of a cold BEC star is a polytrope of index $n=1$, all the physical properties of the star can be derived from the well-known Lane-Emden equation, given by \cite{ChHa}
\begin{equation}\label{laneemden}
\frac{1}{\xi ^{2}}\frac{d}{d\xi }\left (\xi ^{2}\frac{d\theta }{d\xi }\right )=-\theta ,
\end{equation}
where $\theta $ is a dimensionless variable defined via $\rho =\rho
_{c}\theta $, with $\rho _c$ the central density of the star, and $\xi $ is a dimensionless radial coordinate. The general non-singular solution of Eq.~(\ref{laneemden}) is well known, and can be represented  as $\theta \left( \xi \right) =\sin \xi /\xi $. We define the dimensionless radius of the star by the condition $\theta \left( \xi_{1}\right) =0$, giving $\xi _1=\pi $.  Hence the physical radius $R$ of the
Bose-Einstein Condensate star can be represented as a function of the catering length and the fundamental physical constants as \cite{ChHa}
\begin{equation}\label{rxj}
R=\pi \sqrt{\frac{\hbar ^{2}a}{Gm^{3}}}=6.61\left(
\frac{a}{1\;{\rm fm}}\right) ^{1/2}\left(\frac{ m}{2m_{n}}\right) ^{-3/2}{\rm km}.
\end{equation}
For $m=2m_{n}$ and $a=1$ fm the numerical value of the radius of the BEC star is
$R\approx 7$ km. It is interesting to note that the radius of the gravitationally bounded
pure Bose-Einstein Condensate star is independent on the central density and
on the mass of the star, and is fully determined by the physical
characteristics $a$ and $m$ of the condensate only.

The mass of the BEC star can be found  as \cite{ChHa}
\begin{equation}
\label{mrho}
M=4\pi ^{2}\left( \frac{\hbar
^{2}a}{Gm^{3}}\right) ^{3/2}\rho _{c},
\end{equation}
or, equivalently,
\begin{equation}\label{Newtmass}
M=1.84\left( \frac{\rho _{c}}{10^{16}\;{\rm g\;cm^{-3}}}\right) \left(
\frac{a}{1\;{\rm fm}}\right) ^{3/2}\left( \frac{m}{2m_{n}}\right) ^{-9/2}M_{\odot }.
\end{equation}
For $m=2m_{n}$, $a=1\, {\rm fm}$ and $\rho_c=5\times 10^{15}\; {\rm g}/{\rm cm}^3$,  the mass of the condensate is $M\approx 0.92\, M_\odot$. On the other hand by taking for the mass of the condensate particle  the numerical value
$m = m_K ^{*}= m_n/10$, where $m_K ^{*}$ is the kaon mass,   we obtain an extremely high maximum mass
of $M = 63.50 M_{\odot}$ for the kaon condensate star, with a corresponding
radius of $R = 355$ km. Note that the mass of the static condensate can be expressed in terms of the radius and central density by
\begin{equation}
M=\frac{4}{\pi }\rho _{c}R^{3}.
\end{equation}
The above important equation shows that the mean density of the star $\overline{\rho}=3M/4\pi R^{3}$ can be obtained from the central density
of the condensate by the simple relation $\overline{\rho}=3\rho _{c}/\pi ^{2}$.

Under a scaling of the parameters $m$, $a$ and $\rho
_{c}$ so that $m\rightarrow \alpha _{1}m$, $a\rightarrow
\alpha _{2}a$, $\rho _{c}\rightarrow \alpha _{3}\rho _{c}$, the
radii and the masses of the Newtonian BEC stars have the scaling
properties \cite{ChHa},
\begin{equation}\label{scal}
R\rightarrow \alpha _{1}^{-3/2}\alpha _{2}^{1/2}R,M\rightarrow
\alpha _{1}^{-9/2}\alpha _{2}^{3/2}\alpha _{3}M.
\end{equation}

It is important to point out that general relativistic effects impose strong constraints on the structure and global parameters of the BEC stars, and that the values obtained by the simple Newtonian estimates may highly exceed the stability limit imposed by the general relativistic analysis.

\section{Electromagnetic radiation properties of thin accretion disks in
stationary axisymmetric spacetimes}\label{sect2}

In order to investigate the accretion disk properties around neutron, quark and BEC stars we first briefly introduce the general formalism that allows us to describe  the electromagnetic radiation properties of thin accretion disks in stationary axisymmetric spacetimes. In our presentation we closely follow the approach developed in \cite{Ko}.  As a first step in our study we consider the basic kinematic properties of massive particles moving in stable circular orbits in stationary and axially symmetric geometries. The general formalism for obtaining the constants of the motion (energy and angular momentum), the effective gravitational potential and the radii of the marginally stable orbits is presented in some detail. Then we briefly review the basic equations describing the electromagnetic radiation emission from accretion disks, and present the general expressions for the flux, temperature distribution and luminosity of the disk.

\subsection{Massive particle motion in stationary and axially symmetric spacetimes}

In this work our main emphasis is on the analysis of the physical properties and
characteristics of particles that form a thin accretion disk, and  move in circular stable orbits around
general relativistic compact objects. The exterior geometry created by a central dense object is assumed to be stationary and axially
symmetric, given in full generality by the following metric,
\begin{equation}  \label{rotmetr1}
ds^2=g_{tt}\,dt^2+2g_{t\phi}\,dt d\phi+g_{rr}\,dr^2
+g_{\theta\theta}\,d\theta^2+g_{\phi\phi}\,d\phi^2\,.
\end{equation}
In the equatorial approximation,  i.e.,
$|\theta-\pi|\ll 1$, which we adopt in our study, all the metric functions $g_{tt}$, $g_{t\phi}$, $g_{rr}$, $
g_{\theta\theta}$ and $g_{\phi\phi}$ depend on the radial
coordinate $r$ only \cite{PaTh74,Th74}. In the
following we denote the square root of the determinant of the
metric tensor by $\sqrt{-g}$.

In order to determine the electromagnetic properties of the disk  we first obtain
the radial dependence of the angular velocity $\Omega $, of the
specific energy $ \widetilde{E}$ and of the specific angular
momentum $\widetilde{L}$ for particles moving in circular orbits
around compact objects in the geometry given by Eq.~(\ref{rotmetr1}). All these physical parameters can be obtained from the geodesic equations,  which for the considered metric take the following form \cite{na}-\cite{mod}
\begin{eqnarray}
\frac{dt}{ds}&=&\frac{\widetilde{E}
g_{\phi\phi}+\widetilde{L}g_{t\phi}}{
g_{t\phi}^2-g_{tt}g_{\phi\phi}}\,,  \label{geodeqs1} \\
\frac{d\phi}{ds}&=&-\frac{\widetilde{E}
g_{t\phi}+\widetilde{L}g_{tt}
}{g_{t\phi}^2-g_{tt}g_{\phi\phi}}\,,  \label{geodeqs2} \\
g_{rr}\left(\frac{dr}{ds}\right)^2&=&-1+\frac{\widetilde{E}^2
g_{\phi\phi}+2\widetilde{E}\widetilde{L}g_{t\phi}
+\widetilde{L}^2g_{tt}}{ g_{t\phi}^2-g_{tt}g_{\phi\phi}}\,.
\label{geodeqs3}
\end{eqnarray}
We define an effective potential term $V_{eff}(r)$ as
\begin{equation}  \label{roteffpot}
V_{eff}(r)=-1+\frac{\widetilde{E}^2
g_{\phi\phi}+2\widetilde{E}\widetilde{L} g_{t\phi}
+\widetilde{L}^2g_{tt}}{g_{t\phi}^2-g_{tt}g_{\phi\phi}}\,.
\end{equation}

For particles moving in stable circular orbits in the equatorial plane the potential $V_{eff}(r)$ must satisfy the following
two important conditions: $V_{eff}(r)=0$ and $V_{eff,\;r}(r)=0$, respectively, where a coma denotes the derivative with respect to the radial coordinate $r$. From these
conditions we obtain the specific energy, the specific angular
momentum and the angular velocity of particles moving in circular
orbits for the case of spinning general relativistic compact
objects in the form \cite{Ko}
\begin{eqnarray}
\widetilde{E}&=&-\frac{g_{tt}+g_{t\phi}\Omega}{\sqrt{-g_{tt}
-2g_{t\phi}\Omega-g_{\phi\phi}\Omega^2}}\,,  \label{rotE} \\
\widetilde{L}&=&\frac{g_{t\phi}+g_{\phi\phi}\Omega}{\sqrt{-g_{tt}
-2g_{t\phi}\Omega-g_{\phi\phi}\Omega^2}}\,,  \label{rotL} \\
\Omega&=&\frac{d\phi}{dt}=\frac{-g_{t\phi,r}+\sqrt{(g_{t\phi,r})^2
-g_{tt,r}g_{\phi\phi,r}}}{g_{\phi\phi,r}}\,.  \label{rotOmega}
\end{eqnarray}
The marginally stable orbit around the central object can be
determined from the supplementary condition $V_{eff,\;rr}(r)=0$, which provides
the following important relationship
\begin{eqnarray}
\widetilde{E}^{2}g_{\phi\phi,rr}+2\widetilde{E}\widetilde{L}
g_{t\phi,rr}+ \widetilde{L}^{2}g_{tt,rr} -(g_{t\phi}^{2}
-g_{tt}g_{\phi\phi})_{,rr} =0.\nonumber\\  \label{mso-r}
\end{eqnarray}

Once the  metric
coefficients $g_{tt}$, $g_{t\phi}$ and $g_{\phi\phi}$ are explicitly given, by inserting Eqs.~(\ref{rotE})-(\ref{rotOmega}) into Eq.~
(\ref{mso-r}), and numerically solving this equation for the radial coordinate $r$, we obtain the
radii of the marginally stable orbits of massive particles in stable circular orbits around a rotating general relativistic high density compact object \cite{na}-\cite{Ko}.

\subsection{Electromagnetic emissivity of thin accretion disks}

In the following we concentrate on thin accretion disks, that is, accretion disks having their vertical size
negligible as compared to their horizontal extensions. This means that the
disk height $H$, which we define as the maximum half thickness of the disk
in the vertical direction, is much smaller than the
characteristic radius $R$ of the disk, $H \ll R$. We further assume that the thin disk is in
hydrodynamical equilibrium. Moreover, we fully neglect the possible effects of the  pressure gradient, and of the
vertical entropy gradient in the disk.
The heat generated by internal stresses
and dynamical friction is emitted over the disk surface, and this emission
prevents the disk from heating up at extremely high temperatures. On the other hand, this equilibrium determines the disk
to stabilize its thin vertical size. An important astrophysical parameter, the inner
edge of the thin disk, is located at the marginally stable orbit of the central object gravitational
potential. Hence the accreting matter has a Keplerian motion in
higher orbits.

In the steady state accretion disk model, the mass accretion rate
$\dot{M}_{0}$ is a constant that does not change
in time. An effective physical description of the disk properties is obtained by averaging all the quantities describing the orbiting gas
over a characteristic time scale, e.g. $\Delta t$,
over the azimuthal angle $\Delta \phi =2\pi $, and over the height $H$, respectively \cite{ShSu73,
NoTh73,PaTh74}.

As we have already seen, the particles moving in Keplerian orbit around the dense compact object
with  the four-velocity $u^{\mu }$, an have a rotational velocity $\Omega =d\phi /dt$, a specific
energy $\widetilde{E} $, and a specific angular momentum
$\widetilde{L}$. In the steady state thin disk
model all these quantities depend only on the radii $r$ of the orbits. The particles
 form a disk of an
averaged surface density $\Sigma $, which is obtained as the
average of the rest mass density $\rho _{0}$ of the gas in the vertical direction. The
 matter in the disk is described by an anisotropic fluid
source.  The density $\rho _{0}$ of the disk, the energy flow vector
$q^{\mu }$ and the dissipative part of the energy-momentum tensor $t^{\mu \nu }$ are all measured in
the averaged rest-frame. In this specific frame the specific heat and heat transfer processes are neglected. Then,
one important parameter characterizing the disk structure is the surface density of
the disk \cite{NoTh73,PaTh74},
\begin{equation}
\Sigma(r) = \int^H_{-H}\langle\rho_0\rangle dz,
\end{equation}
which is obtained as the averaged rest mass density $\langle\rho_0\rangle$ over
$\Delta t$ and $ 2\pi$. Another important disk parameter is the torque
\begin{equation}
W_{\phi}{}^{r} =\int^H_{-H}\langle t_{\phi}{}^{r}\rangle dz,
\end{equation}
representing the average of the component $\langle t^r_{\phi} \rangle$ of the energy-momentum tensor over
$\Delta t$ and $2\pi$. The time and orbital average of the energy
flux vector gives the radiation flux $F(r)$ from the
disk surface as
\be
F(r)=\langle q^z \rangle.
\ee

The energy-momentum tensor of the matter in the disk is represented in its standard form  according to
\begin{equation}
T^{\mu \nu }=\rho_{0}u^{\mu }u^{\nu }+2u^{(\mu }q^{\nu )}+t^{\mu \nu },
\end{equation}
where the four-velocity of the matter satisfies the conditions $u_{\mu }q^{\mu }=0$, $u_{\mu }t^{\mu \nu }=0$. The
four-vectors of the energy and angular momentum flux are defined
by $-E^{\mu }\equiv T_{\nu }^{\mu }{}(\partial /\partial t)^{\nu
}$ and $J^{\mu }\equiv T_{\nu }^{\mu }{}(\partial /\partial \phi
)^{\nu }$, respectively. Note that the structure equations of the thin disk
can be derived by integrating the conservation laws of the rest
mass, of the energy, and of the angular momentum of the gas forming the disk,
respectively \cite{NoTh73,PaTh74}. From the equation of the rest
mass conservation, $\nabla _{\mu }(\rho _{0}u^{\mu })=0$, it
follows first the important result that the time averaged mass accretion rate is independent of the disk radius,
\begin{equation}
\dot{M_{0}}\equiv -2\pi \sqrt{-g}\Sigma u^{r}=\mathrm{constant}.
\label{conslawofM}
\end{equation}

The conservation law $\nabla _{\mu }E^{\mu }=0$ of the energy can be reformulated in
the integral form, more interesting from a physical point of view, as \cite{Ko}
\begin{equation}
\lbrack \dot{M}_{0}\widetilde{E}-2\pi \sqrt{-g}\Omega W_{\phi
}{}^{r}]_{,r}=4\pi \sqrt{-g}F \widetilde{E}.  \label{conslawofE}
\end{equation}%
The above equation shows  that the energy transported by the mass flow,
$\dot{M}_{0} \widetilde{E}$, and the energy transported by the
dynamical stresses in the disk, $2\pi \sqrt{-g}\Omega W_{\phi }{}^{r}$, is
exactly compensated by the energy radiated away from the surface of the
disk, $4\pi \sqrt{-g}F\widetilde{E}$. The law of the angular momentum
conservation, $\nabla _{\mu }J^{\mu }=0$, indicates the equilibrium between
the three forms of the angular momentum transport,
\begin{equation}
\lbrack \dot{M}_{0}\widetilde{L}-2\pi rW_{\phi }{}^{r}]_{,r}=4\pi
\sqrt{-g}F \widetilde{L}\;\;.  \label{conslawofL}
\end{equation}

By eliminating $W_{\phi }{}^{r}$ from Eqs.~(\ref{conslawofE}) and
(\ref {conslawofL}), with the use of the universal energy-angular
momentum relation $ dE=\Omega dJ$ for circular geodesic orbits, written in
the form $\widetilde{E} _{,r}=\Omega \widetilde{L}_{,r}$, the flux
$F$ of the electromagnetic energy emitted by the disk surface is obtained  as \cite{NoTh73,PaTh74, Ko},
\begin{equation}
F(r)=-\frac{\dot{M}_{0}}{4\pi \sqrt{-g}}\frac{\Omega
_{,r}}{(\widetilde{E}-\Omega
\widetilde{L})^{2}}\int_{r_{in}}^{r}(\widetilde{E}-\Omega
\widetilde{L}) \widetilde{L}_{,r}dr\;\;,  \label{F}
\end{equation}
where $r_{in}$ is the inner edge of the disk. In our study we assume that $r_{in}=r_{ms}$. Note that the flux depends
of the specific energy, angular momentum and  angular
velocity of the gas motion around the central general relativistic compact object.

The gas forming the accretion disk in the steady-state thin disk model is
assumed to be in thermodynamical equilibrium. Therefore the
radiation emitted by the disk surface may be considered as a
perfect black body radiation, depending on the temperature only, with the energy flux given by
\be\label{T}
F(r)=\sigma _{SB}T^{4}(r),
\ee
where $\sigma _{SB}$ is the Stefan-Boltzmann
constant. Once the radiative flux is known, from Eq.~(\ref{T}) we obtain the temperature distribution on the disk surface.

The observed luminosity $L\left( \nu \right)$ of the disk surface  has a
redshifted black body spectrum, given by \cite{To02},
\bea\label{L}
L\left( \nu \right) &=&4\pi d^{2}I\left( \nu \right) =\frac{8\pi  h\cos \gamma }{c^2}\times \nonumber\\
&&\int_{r_{in}}^{r_{f}}\int_0^{2\pi}\frac{\nu^{3}_e r d\phi dr }{\exp
\left(h\nu_e/k_BT\right) -1},
\eea

In Eq.~(\ref{L}) $h$ is Planck's constant, $k_B$ is Boltzmmann's constant, $d$ is the distance to the source (disk), $I(\nu )$ is the Planck
distribution function, $\nu _e$ is the frequency of the emitted radiation, $\gamma $ is the disk inclination angle,
and $r_{in}$ and $r_{f}$ indicate the positions of the inner and
outer edge of the disk, respectively. In the natural system of units with $\hbar =c=k_B=1$, we obtain for the disk luminosity the expression
\be
L\left( \nu \right) = 16\pi ^2\cos \gamma \int_{r_{in}}^{r_{f}}\int_0^{2\pi}\frac{\nu^{3}_e r d\phi dr }{\exp
\left(2\pi \nu_e/T\right) -1},
\ee

In the following we take the upper limit of integration in Eq.~(\ref{L}) as infinity, that is, we assume
$r_{f}\rightarrow \infty $. Moreover, we expect that the flux from the disk
surface vanishes at $r\rightarrow \infty $. This condition is independent on the geometry of the general
relativistic compact object. The frequency of the emitted  radiation is
given by the relation $\nu_e=\nu(1+z)$, where the redshift factor can be
written as
\begin{equation}\label{rs}
1+z=\frac{1+\Omega r \sin \phi \sin \gamma }{\sqrt{ -g_{tt}-2
\Omega g_{t\phi} - \Omega^2 g_{\phi\phi}}}.
\end{equation}
In Eq.~(\ref{rs}) we have neglected the light bending \cite{Lu79,BMT01}. This approximation  works well for small inclination angles, but it is not as good for large inclination angles (edge-on disks).

Another important characteristics of the mass accretion processes is
the efficiency $\epsilon $ with which the central object converts the mass of the gas
into radiation. The efficiency $\epsilon$ is defined as the ratio of
the rate of the energy of photons emitted from the
disk surface, and the rate at which mass-energy is
transported to the central high density general relativistic object  \cite{NoTh73,PaTh74}.  Both energies are measured at infinity. If all  photons escape to infinity, the efficiency depends
on the specific energy measured at the marginally stable orbit $
r_{ms}$ only,
\begin{equation}
\epsilon =1-\widetilde{E}_{ms}.  \label{epsilon}
\end{equation}

For Schwarzschild black holes the efficiency $\epsilon $ is about
$6\%$. This result is independent on whether the photon capture by the black hole is considered,
or not. For rapidly rotating black holes for which the capture of radiation by the black hole is ignored, $\epsilon $is equal to $42\%$.  For a Kerr black hole with the photon capture explicitly considered
the efficiency is $40\%$  \cite{Th74}.

Note that the fluxes and the emission spectra of the accretion disks around compact objects
satisfy some simple, but important, scaling relations. Such scaling relations can be found when one considers the  scaling
transformation of the radial coordinate, given by $r\rightarrow \widetilde{r}=r/M$,
where $M$ is the mass of the central object. Under such a scaling transformation in general the metric tensor
coefficients are invariant, while the
specific energy, the angular momentum and the angular velocity transform as $%
\widetilde{E}\rightarrow \widetilde{E}$, $\widetilde{L}\rightarrow M\widetilde{L}$ and $%
\Omega \rightarrow \widetilde{\Omega}/M$, respectively. The flux scales as $F(r)\rightarrow F(%
\widetilde{r})/M^{4}$, giving the simple scaling law of the temperature as $%
T(r)\rightarrow T\left( \widetilde{r}\right) /M$. Another important rescaling is the transformation of the frequency
of the emitted radiation as  $\nu \rightarrow \widetilde{\nu}=\nu /M$. Under this rescaling of the radiation frequency
the luminosity of the disk scales as $L\left( \nu \right) \rightarrow
L\left( \widetilde{\nu}\right) /M$ \cite{Ko}. On the other hand, since the flux is proportional
to the accretion rate $\dot{M}_{0}$, we obtain the important result that an increase in the
accretion rate leads to a linear increase of the radiation flux
from the disk.

From an observational point of view, the use of the thermal radiation from thin disks to test the nature of a compact general relativistic object is often called the continuum-fitting method, and was proposed in \cite{Zhang}  to investigate the observational consequences of black hole spin in X-ray binaries. The standard thin accretion disk model introduced in \cite{PaTh74} was used to analyze the emissivity of accretion disks around black holes.
In the case of Kerr black holes the observed disk spectrum, however, bears several important corrections to the
simple formula given by Eq.~(\ref{F}). Since the X-rays are emitted on the hot inner disk, in this region electron
scattering may dominate over the free-free absorption. As a consequence the color temperature
may be greater than the effective temperature, with the inner disk radiating like a diluted black body. The general relativistic corrections near the event horizon (gravitational redshift and focusing) can cause the observed color temperature and the integrated flux to be different from the true local values. The continuum-fitting method was extensively used to observationally study the spin of black holes \cite{cont}.

 \section{Equations of state and physical parameters of the neutron, quark and BEC stars}\label{sect3}

To set the stage for our study, in the present Section we introduce first the equations of state of the dense nuclear matter considered in our comparative study of the neutron, quark and Bose-Einstein Condensate stars, and present some of the basic dynamical properties of the corresponding rotating stellar models. In order to compare the properties of the BEC stars with other compact objects we choose five equations of state of neutron matter, and the bag model equation of state for quark matter. We restrict our study to rotating BEC stars satisfying the $n=1$ polytropic equation of state. The astrophysical parameters of the stars described by the adopted equations of state are also presented in detail. In order to compare the properties of the stellar models we consider three classes of models. In the first model we fix the values of the mass and of the angular velocity for all stars. As a second case we consider compact stars rotating at the maximum (Keplerian) frequencies. And finally, as a third class of models we consider stars having the same central density and fixed ratio of the equatorial and polar radii, respectively.

\subsection{Equations of state of the neutron, quark and BEC matter}

In order to obtain a consistent and realistic physical description of the rapidly rotating general relativistic compact stars, as a first step we have to adopt the equations of state for the dense neutron, quark,  and Bose-Einstein Condensate stellar matter, respectively. In the present comparative study of the physical properties of the accretion disks onto rapidly rotating compact general relativistic objects  we adopt the following equations of state \cite{Ko}:

1) Akmal-Pandharipande-Ravenhall  (APR) EOS \cite{Ak98}. As a first example of an equation of state of dense neutron matter we consider EOS APR, which was obtained numerically by using the variational chain summation methods and the  Argonne $v_{18}$ two-nucleon interaction, respectively. Note that boost corrections to the two-nucleon interaction, giving the leading relativistic effect of order $(v/c)^2$, as well as three-nucleon interactions, are also included in the basic nuclear Hamiltonian.  The dense matter density range described by EOS APR is from $2\times 10^{14}$ g/cm$^3$ to $2.6\times 10^{15}$ g/cm$^3$. The maximum mass limit in the static case for this EOS is $2.20 M_{\odot}$.  To obtain a full description of the neutron star properties the APR EOS is joined to the composite Baym-Bethe-Pethick (BBP) ($\varepsilon /c^2>4.3\times10^{11}$g/cm$^3$) \cite{Ba71a} - Baym-Pethick-Sutherland (BPS) ($10^4$ g/cm$^3$ $<4.3\times 10^{11}$g/cm$^3$) \cite{Ba71b} - Feynman-Metropolis-Teller (FMT) ($\varepsilon/c^2<10^4$ g/cm$^3$) \cite{Fe49} equations of state, respectively.

2) Douchin-Haensel (DH) EOS \cite{DoHa01}. EOS DH represents a complete equation of state of the neutron star matter. It describes  both the neutron star crust as well as its liquid core. It is constructed by using the effective nuclear interaction SLy of the Skyrme type. The corresponding interaction potential describes very well the properties of very rich neutron matter. On the other hand the structure of the crust, and its EOS, is obtained in the zero temperature approximation only, and under the assumption of the ground state composition. The EOS of the liquid core is calculated assuming (minimal) $npe\mu $ composition. The density range is from $3.49\times 10^{11}$ g/cm$^3$ to $4.04\times 10^{15}$ g/cm$^3$. The minimum and maximum masses of the static neutron stars for the DH EOS are $0.094M_{\odot}$ and $2.05 M_{\odot}$, respectively.

3) 	Shen-Toki-Oyamatsu-Sumiyoshi (STOS) EOS \cite{Shen}. The STOS equation of state of nuclear matter is obtained by using the relativistic mean field theory with nonlinear $\sigma $ and $\omega $ terms. It can be used in a wide neutron matter density and temperature range, with various proton fractions. The EOS was specifically designed for the use of supernova explosion simulation and for the calculations of the neutron star properties. To compute the properties of inhomogeneous nuclear matter, where heavy nuclei are formed together with free nucleon gas, the  Thomas-Fermi approximation is used. In the present paper we consider only the zero temperature STOS EOS, namely the EOS for $T=0$, and we denote it as STOS0.

4) Relativistic Mean Field (RMF) equations of state with isovector scalar mean field in the presence  of the $\delta $-meson field- RMF soft and RMF stiff EOSs \cite{Kubis}.  Even that in symmetric nuclear matter the $\delta $-meson mean field vanishes, this field still can significantly influence the properties of asymmetric nuclear matter in neutron stars. Note that the Relativistic Mean Field contribution due to the $\delta $-field to the nuclear symmetry energy is smaller than zero. Therefore the energy per particle of neutron matter is larger at high densities, as compared to the case when no $\delta $-field included. Also, the proton fraction of $\beta $-stable matter increases. Splitting of proton and neutron effective masses due to the $\delta $-field can significantly affect transport properties of neutron star matter. These equations of state can be parameterized by the coupling parameters $C_{\sigma }^2=g_{\sigma }^2/m_{\sigma }^2$, $C_{\omega }^2=g_{\omega }^2/m_{\omega }^2$, $\bar{b}=b/g_{\sigma}^3$ and $\bar{c}=c/g_{\sigma}^4$, respectively, where $m_{\sigma }$ and $m_{\omega }$ are the masses of the respective mesons, and $b$ and $c$ are the coefficients in the potential energy $U\left(\sigma \right)$ of the $\sigma $-field. The soft RMF EOS is parameterized by $C_{\sigma }^2=1.582$ fm$^2$, $C_{\omega }^2=1.019$ fm$^2$, $\bar{b}=-0.7188$ and $\bar{c}=6.563$, while the stiff RMF EOS is parameterized by $C_{\sigma }^2=11.25$ fm$^2$, $C_{\omega }^2=6.483$ fm$^2$, $\bar{b}=0.003825$ and $\bar{c}=3.5\times 10^{-6}$, respectively.

5) Baldo-Bombaci-Burgio (BBB) EOS \cite{baldo}. The BBB EOS is an EOS for asymmetric nuclear matter. It is derived from the Brueckner-Bethe-Goldstone many-body theory with explicit three-body forces taken into account. Two EOS's are obtained, one corresponding to the Argonne AV14 (BBBAV14), and the other to the Paris two-body nuclear force (BBBParis), implemented by the Urbana model for the three-body force. The maximum static mass configurations are $M_{max} = 1.8 M_{\odot}$ and  $M_{max} = 1.94 M_{\odot}$ when the AV14 and Paris interactions are used, respectively.  The onset of direct Urca processes occurs at particle number densities $n\geq 0.65$ fm$^{-3}$ for the AV14 potential, and $n\geq 0.54$ fm$^{-3}$ for the Paris potential. The comparison with other microscopic models for the EOS shows noticeable differences, which can also influence significantly the neutron star properties. The density range for this EOS is from $1.35\times 10^{14}$ g/cm$^3$ to $3.507\times 10^{15}$ g/cm$^3$.

6) Bag model equation of state for quark matter - Q EOS \cite{Wi84}. For the description of the quark matter we adopt  a simple phenomenological description, based on the MIT bag model equation of state. Hence we assume that in quark matter the  pressure $p$ is related to the energy density $\rho $ by
\begin{equation}
p=\frac{1}{3}\left(\rho-4B\right)c^2,
\end{equation}
where the parameter $B$, called the bag constant, is the difference between the energy density of the perturbative and
non-perturbative Quantum Chromodynamic vacuum. For the bag constant we adopt the numerical value $B=4.2\times 10^{14}$ g/cm$^3$ \cite{Wi84}.

7) The Bose-Einstein Condensate equation of state (BEC) EOS. For the Bose-Einstein Condensate stars we adopt the $n=1$ polytropic equation of state, given by
\begin{equation}
P\left( \rho \right) =K\rho ^{2},
\end{equation}
with
\begin{equation}
K=\frac{2\pi \hbar ^{2}a}{m^{3}} =0.1856\times 10^5 \left(\frac{a}{1\;{\rm fm}}\right)\left(\frac{m}{2m_n}\right)^{-3},
\end{equation}
where $m_n=1.6749\times 10^{-24}$ g is the mass of the neutron. The behavior of the BEC EOS, as well as of the corresponding stellar models essentially depends on the ratio $a/m^3$ of the scattering length and of the condensate particle mass. In the present study we restrict our analysis to three values of the ratio $\left(a/1\;{\rm fm}\right)\left(m/2m_n\right)^{-3}$: $\left(a/1\;{\rm fm}\right)\left(m/2m_n\right)^{-3}=15$, $\left(a/1\;{\rm fm}\right)\left(m/2m_n\right)^{-3}=30$, and $\left(a/1\;{\rm fm}\right)\left(m/2m_n\right)^{-3}=50$, respectively. We denote the corresponding equations of state of the BEC matter by BEC15, BEC30, and BEC50, respectively.

\begin{figure}[tbh]
\centering
\includegraphics[width=9cm]{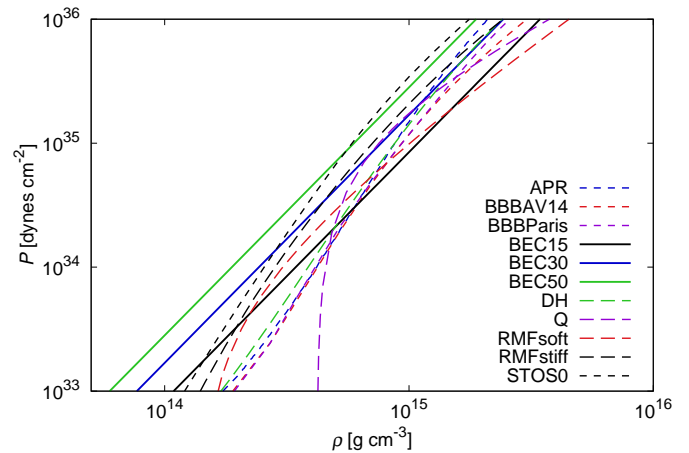}
\caption{Variation of the pressure as a function of the density for the considered equations of state of the neutron, quark and BEC dense matter.}\label{fig1}
\end{figure}

The variation of the considered equations of state with respect to the density of the dense stellar matter is represented in Fig.~\ref{fig1}.

The low density behavior of the BEC15 and BEC30 EOSs is significantly different as compared to the other considered equations of state, indicating the possibility of the existence of stable BEC stars at pressures lower than the neutron matter pressure. In the adopted range of densities the BEC EOSs show a linear pressure-density dependence.

\subsection{Astrophysical parameters of the neutron, quark and BEC stars}

In quasi-isotropic coordinates the metric outside a rotating compact general relativistic star can be represented as \cite{Ster}
\bea
ds^2&=&-e^{\bar{\gamma }+\bar{\rho }}dt^2+e^{2\bar{\alpha } }\left(d\bar{r}^2+\bar{r}^2d\theta ^2\right)+e^{\bar{\gamma }-\bar{\rho }}\bar{r}^2\sin ^2\theta \times \nonumber\\
&&\left(d\phi -\bar{\omega}dt\right)^2,
\eea
where the metric potentials $\bar{\gamma }$, $\bar{\rho }$, $\bar{\alpha }$ and the angular velocity of the stellar fluid  $\bar{\omega}$,  measured relative to the local inertial frame, are all functions of the quasi-isotropic radial coordinate $\bar{r}$, and of the polar angle $\theta $. The RNS code computes numerically the metric functions in a quasi-spheroidal coordinate system, as functions of the parameter $s=\bar{r}/\left(\bar{r}+\bar{r}_e\right)$, where $\bar{r}_e$ is the equatorial radius of the star, which we have converted into Schwarzschild-type coordinates $r$ according to the equation $r=\bar{r}\exp\left[\left(\bar{\gamma }-\bar{\rho }\right)/2\right]$. To obtain the radius of the marginally (or innermost) stable  circular orbits $r_{ms}$ we use a truncated form of the analytical approximation given as \cite{ShSa98},
\begin{eqnarray}
\frac{r_{ms}}{6M}&\approx &1-0.54433q-0.22619q^2+0.17989Q_2-\nonumber\\
&&0.23002q^3+
  0.26296qQ_2-0.29693q^4+\nonumber\\
 && 0.44546q^2Q_2,
\end{eqnarray}
where $q=J/M^2$ and $Q_2=-M_2/M^3$, respectively, and where $J$ is the spin angular momentum, and $M_2$ is the quadrupole moment.

\subsubsection{Mass-radius relation for neutron, quark and BEC stars}

\begin{figure*}[tbp]
\centering
\includegraphics[width=8cm]{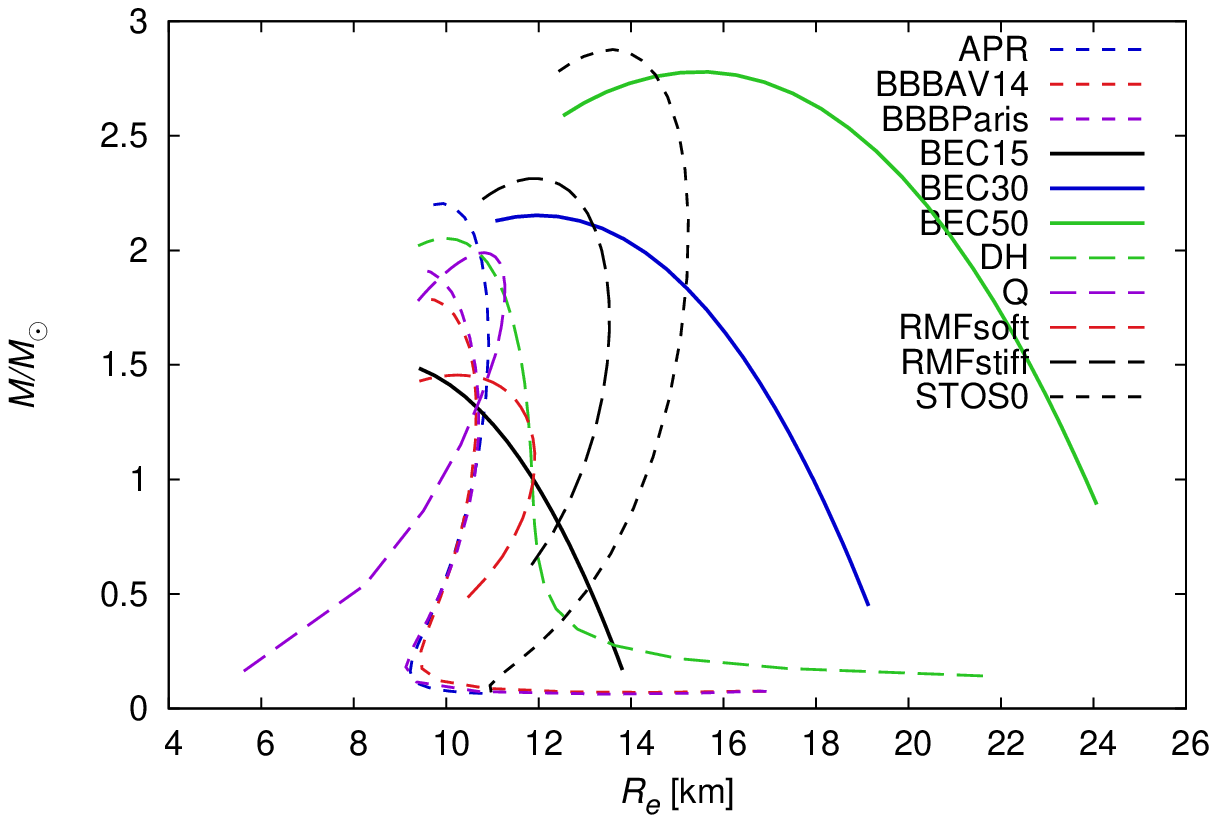}
\includegraphics[width=8cm]{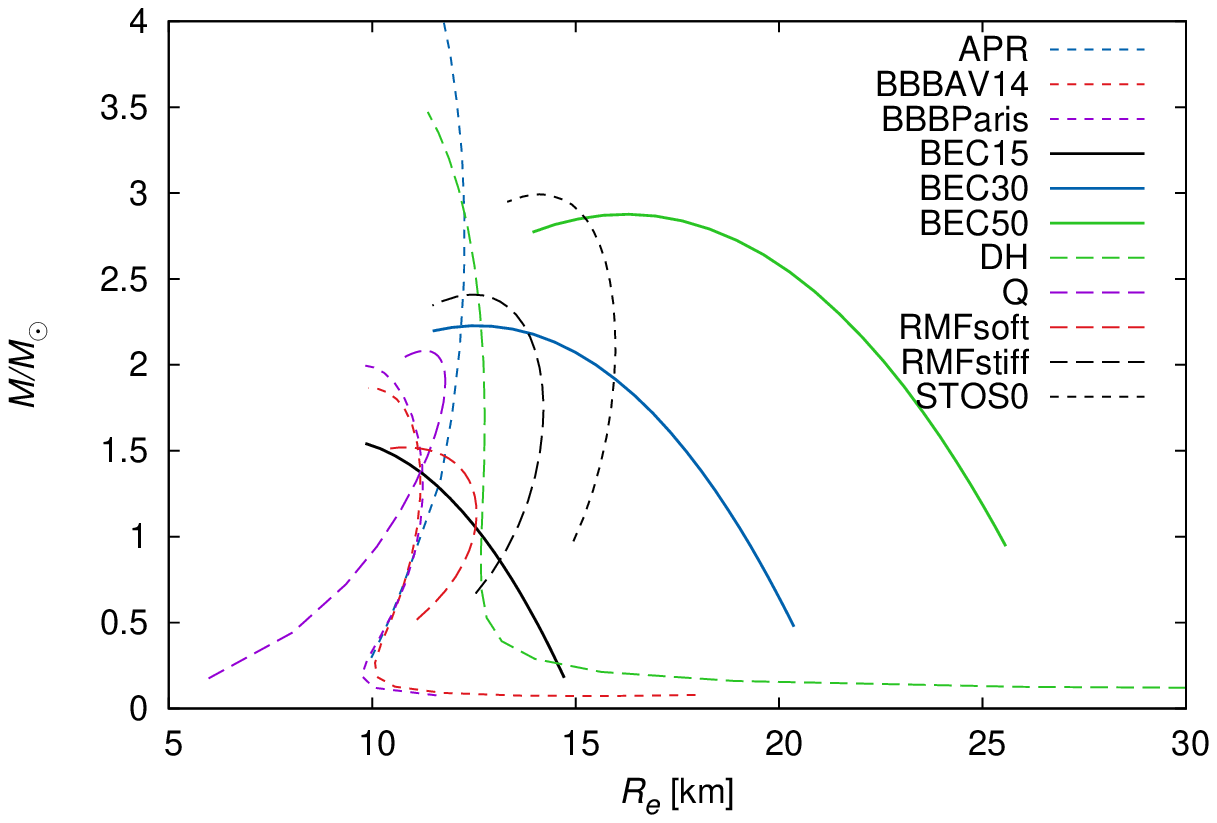}
\caption{Mass-circumferential radius relation for the rotating neutron, quark and BEC stars for $r_p/r_e=1$ (left figure) and $r_p/r_e=0.9$ (right figure).}
\label{fig2}
\end{figure*}

\begin{figure*}[tbp]
\centering
\includegraphics[width=8cm]{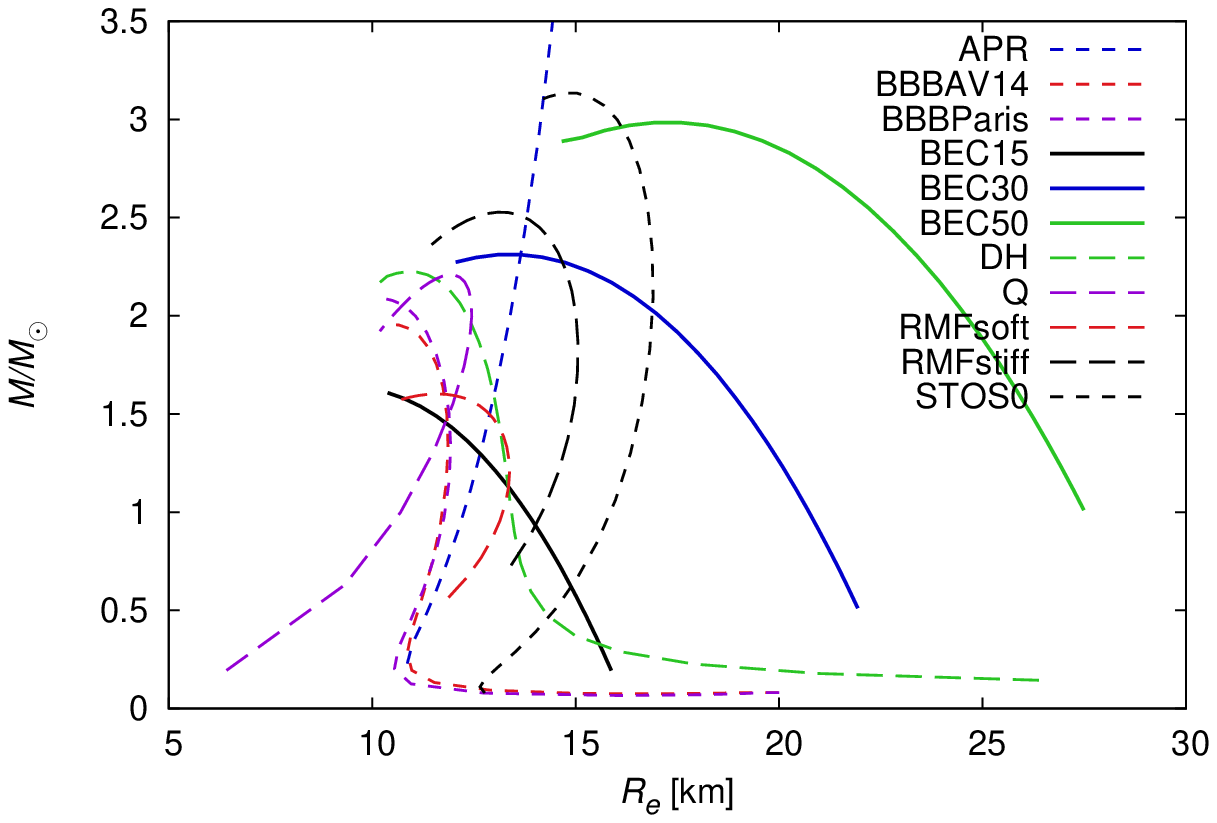}
\includegraphics[width=8cm]{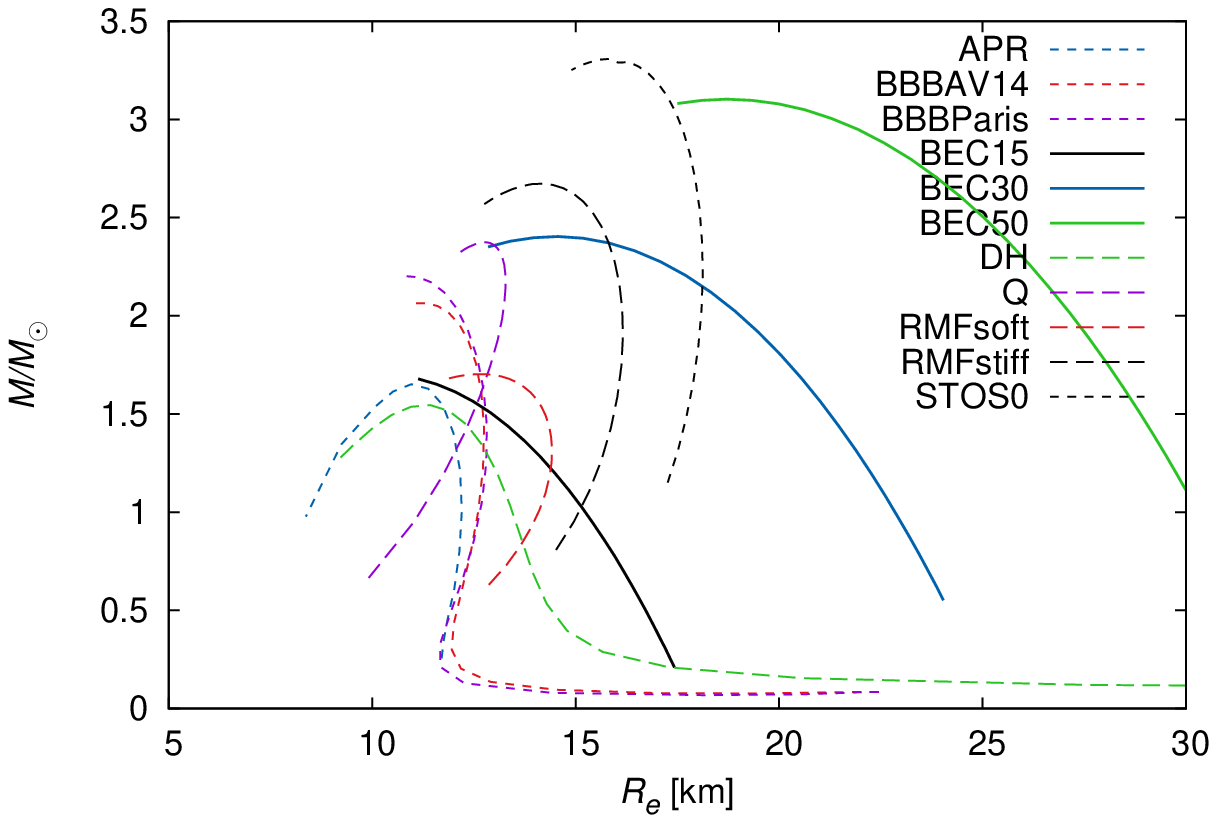}
\caption{Mass-circumferential radius relation for the rotating neutron, quark and BEC stars for $r_p/r_e=0.8$ (left figure) and $r_p/r_e=0.7$ (right figure).}
\label{fig3}
\end{figure*}

The mass-radius relation of the compact general objects for the considered equations of state of the dense nuclear matter, $M=M\left(R_e\right)$, where $R_e$ is the circumferential radius at
the equator, are presented, for four fixed values of the ratio $r_p/r_e$, where $r_p$ is the polar radius, and $r_e$ is the equatorial radius of the star, in Figs.~\ref{fig2} and \ref{fig3}, respectively.

    As one can see from the Figures, the BEC stars form a distinct class of stellar objects, as compared to the group of "standard" neutron and quark stars. The mass-radius relation is systematically shifted to the right region of the $M=M\left(R_e\right)$ relation, has a specific shape, and indicates a much larger radius for the BEC star. The radius increases with increasing $a/m^3$, so that for $\left(a/1\;{\rm fm}\right)\left(m/2m_n\right)^{-3}=50$ and $r_p/r_e=0.7$ the radius of the maximum mass stable BEC star is of the order of 30 km. Thus a first distinctive signature of rapidly rotating BEC stars is their bigger radius, as compared to the "standard" neutron and quark stars, indicating a large value of the coefficient $a/m^3$. On the other hand, for $\left(a/1\;{\rm fm}\right)\left(m/2m_n\right)^{-3}<20$, the radius of the maximum mass stable BEC star is of the same order as the radii of the "standard" stars, $R_e\approx 8-15$ km. Significant differences appear in the maximum masses of the stars, with the maximum allowable mass of the BEC50 star varying between $2.8M_{\odot}$ (static case) and $3.0M_{\odot}$, for the rapidly rotating star with $r_p/r_e=0.7$. For the BEC15 star, the maximum  mass of the stable configuration ranges between $1.5M_{\odot}$ (static case), and $1.6M_{\odot}$, for $r_p/r_e=0.7$. Hence a specific mass-radius relation provides a first distinctive feature of the Bose-Einstein Condensate stars, as compared to the considered neutron and quark stars.

\subsubsection{Models with fixed mass and angular velocity}

In analyzing the emissivity properties of the accretion disks we consider three types of stellar models, whose main astrophysical properties are presented in the following in a tabular form. In all Tables, $\rho _c$ is the central density, $M$ is the gravitational mass, $M_0$ is the rest mass, $R_e$ is the circumferential radius at
the equator, $\Omega $ is the angular velocity, $\Omega _p$ is the angular velocity of a particle in circular orbit at the equator, $T/W$ is the rotational-gravitational energy ratio, $cJ/GM_{\odot}^2$ is the angular momentum, $I$ is the moment of inertia, $\Phi_2$ gives the mass quadrupole moment $M_2$ so that $M_2=c^4\Phi_2/(G^2M^3M^3_{\odot})$, $h_+$ is the height from the surface of the last stable co-rotating circular orbit in the equatorial plane, $h_{-}$ is the height from surface of the last stable counter-rotating circular orbit in the equatorial plane, $\omega _c/\Omega$ is the ratio of the central value of the potential $\omega $ to $\Omega $, $r_e$ is the coordinate equatorial radius, and $r_p/r_e$ is the axes ratio (polar to equatorial), respectively.

The physical properties of the neutron, quark and BEC stars with fixed mass, $M\approx 1.8M_{\odot}$ and angular velocity $\Omega \approx 5\times 10^3$ s$^{-1}$ are presented in Table~\ref{table1}.

\begin{widetext}
\begin{center}
\begin{table}
	  \scalebox{1} {
	   	\begin{tabular}{|l| c| c| c| c| c| c| c| c| c| c| c| c|}
	   	  \hline \hline
	   	  EOS & APR & BBBAV14 & BBBParis & BEC15 & BEC30 & BEC50 & DH & Q & RMFsoft & RMFstiff & STOS0 \\ \hline
	   	  $\rho_c [10^{15}$ g/cm$^3]$ & 1.20 & 2.15 & 1.70 & 5.00 & 0.50 & 0.19 & 1.29 & 0.895 & 2.00 & 0.57 & 0.369 \\
	   	  $M [M_\odot]$ & 1.798 & 1.805 & 1.799 & 1.541 & 1.811 & 1.8 & 1.805 & 1.806 & 1.525 & 1.803 & 1.854 \\
	   	  $M_0 [M_\odot]$ & 2.083 & 2.108 & 2.095 & 1.735 & 1.944 & 1.895 & 2.054 & 1.953 & 1.705 & 2.003 & 2.016 \\
	   	  $R_e$ [km] & 11.437 & 10.35 & 10.706 & 8.35 & 24.917 & 30.661 & 12.02 & 11.979 & 11.344 & 15.812 & 21.038 \\
	   	  $\Omega [10^3$ s$^{-1}]$ & 5.008 & 5.034 & 5.039 & 5.04 & 4.269 & 2.581 & 5.008 & 5.006 & 4.999 & 5.009 & 4.902 \\
	   	  $\Omega_p [10^3$ s$^{-1}]$ & 12.473 & 14.414 & 13.705 & 18.341 & 3.983 & 2.961 & 11.593 & 11.781 & 11.674 & 7.972 & 5.374 \\
	   	  $T / W [10^{-2}]$ & 3.449 & 2.375 & 2.712 & 1.113 & 9.756 & 9.429 & 3.46 & 4.836 & 3.312 & 8.981 & 14.911 \\
	   	  $cJ/GM^2_\odot$ & 1.142 & 0.953 & 1.01 & 0.487 & 2.233 & 2.418 & 1.155 & 1.389 & 0.826 & 2.011 & 2.953 \\
	   	  $I [10^{45}$ g cm$^2]$ & 2.005 & 1.663 & 1.761 & 0.849 & 4.596 & 8.233 & 2.028 & 2.438 & 1.452 & 3.528 & 5.294 \\
	   	  $\Phi_2 [10^{43}$ g cm$^2]$ & 8.263 & 4.325 & 5.365 & 0.978 & 69.829 & 143.288 & 8.433 & 14.988 & 6.717 & 43.932 & 106.248 \\
	   	  $h_+$ [km] & 0.00 & 3.39 & 2.93 & 3.82 & 0.00 & 0.00 & 1.65 & 0.00 & 0.00 & 0.00 & 0.00 \\
	   	  $h_-$ [km] & 7.958 & 8.352 & 8.166 & 6.845 & 1.296 & 0.00 & 7.47 & 0.00 & 0.00 & 0.00 & 7.859 \\
	   	  $\omega_c/\Omega [10^{-1}]$ & 5.809 & 6.667 & 6.343 & 7.769 & 4.159 & 3.025 & 5.856 & 5.252 & 5.519 & 4.526 & 4.103 \\
	   	  $r_e$ [km] & 8.521 & 7.408 & 7.781 & 5.841 & 22.087 & 27.848 & 9.11 & 9.024 & 8.909 & 12.872 & 18.013 \\
	   	  $r_p/r_e$ & 0.894 & 0.925 & 0.915 & 0.958 & 0.54 & 0.623 & 0.885 & 0.86 & 0.883 & 0.728 & 0.54 \\ \hline
	   	\end{tabular}}
\caption{Astrophysical parameters of the neutron, quark and BEC stars with mass $M\approx 1.8M_\odot$, rotating at an angular velocity  $\Omega \approx 5\times10^3$ s$^{-1}$. }\label{table1}
\end{table}
\end{center}
\end{widetext}
The BEC15 EOS does not allow stellar masses of the order of $1.8M_{\odot}$. The mass corresponding to an angular velocity of $\Omega \approx 5\times 10^3$ s$^{-1}$ is $1.541M_{\odot}$, with a small equatorial radius of the order of 6 km, and a high central density $\rho _c=5\times 10^{15}$ g/cm$^3$. Configurations with $1.8M_{\odot}$ can be obtained for the BEC30 and BEC50 EOSs. They have  equatorial radii of the order of 22 and 28 km. A similar, but still smaller equatorial radius $r_e=18$ km, can be found only for the STOS0 EOS. The BEC30, BEC50 and STOS0 configurations have similar central densities. The BEC30 and BEC50 stars have the highest moment of inertia $I$, and highest angular momentum per unit $cJ/GM$, with EOS STOS0 having the closest values of these parameters.

\subsubsection{Models rotating at Keplerian frequencies}

As a second astrophysical model we consider the case of the neutron, quark and BEC stars rotating at Keplerian frequencies. The physical properties of this class of stars are presented in Table~\ref{table2}.

\begin{widetext}
\begin{center}
 \begin{table}
      \scalebox{1} {
    	\begin{tabular}{| l| c| c| c| c| c| c| c| c| c| c| c| c| }
    	  \hline \hline	
    	  EOS & APR & BBBAV14 & BBBParis & BEC15 & BEC30 & BEC50 & DH & Q & RMFsoft & RMFstiff & STOS0 \\ \hline
    	  $\rho_c [10^{15}$ g/cm$^3]$ & 1.00 & 1.00 & 1.00 & 1.00 & 1.00 & 1.00 & 1.00 & 1.00 & 1.00 & 1.00 & 1.00 \\
    	  $M [M_\odot]$ & 1.93 & 1.646 & 1.671 & 1.286 & 2.311 & 3.18 & 1.784 & 2.807 & 1.841 & 2.786 & 3.492 \\
    	  $M_0 [M_\odot]$ & 2.183 & 1.83 & 1.86 & 1.381 & 2.557 & 3.567 & 1.982 & 3.316 & 2.039 & 3.229 & 4.147 \\
    	  $R_e$ [km] & 15.493 & 15.279 & 15.408 & 16.831 & 20.132 & 22.291 & 16.249 & 17.152 & 17.793 & 18.435 & 19.148 \\
    	  $\Omega [10^3$ s$^{-1}]$ & 8.398 & 7.948 & 7.919 & 6.07 & 6.105 & 6.059 & 7.439 & 9.031 & 6.912 & 7.746 & 7.991 \\
    	  $\Omega_p [10^3$ s$^{-1}]$ & 8.405 & 7.948 & 7.919 & 6.07 & 6.105 & 6.06 & 7.439 & 9.031 & 6.912 & 7.746 & 7.991 \\
    	  $T / W [10^{-2}]$ & 15.064 & 14.074 & 14.347 & 9.932 & 9.71 & 9.458 & 11.49 & 21.989 & 16.232 & 15.493 & 14.885 \\
    	  $cJ/GM^2_\odot$ & 2.88 & 2.081 & 2.168 & 1.138 & 3.295 & 6.038 & 2.154 & 7.183 & 2.913 & 5.948 & 8.975 \\
    	  $I [10^{45}$ g cm$^2]$ & 3.014 & 2.301 & 2.406 & 1.648 & 4.744 & 8.757 & 2.545 & 6.99 & 3.704 & 6.748 & 9.871 \\
    	  $\Phi_2 [10^{43}$ g cm$^2]$ & 47.829 & 37.726 & 40.021 & 25.417 & 54.725 & 84.065 & 33.907 & 130.893 & 73.347 & 99.348 & 123.857 \\
    	  $h_+$ [km] & 0.00 & 0.00 & 0.00 & 0.00 & 0.00 & 0.00 & 0.00 & 0.00 & 0.00 & 0.87 & 1.63 \\
    	  $h_-$ [km] & 10.41 & 7.397 & 7.695 & 1.893 & 8.805 & 15.843 & 6.918 & 21.205 & 9.532 & 17.732 & 24.495 \\
    	  $\omega_c/\Omega [10^{-1}]$ & 5.789 & 5.176 & 5.203 & 4.171 & 6.047 & 7.332 & 5.439 & 6.978 & 5.154 & 6.905 & 7.945 \\
    	  $r_e$ [km] & 12.29 & 12.597 & 12.679 & 14.81 & 16.422 & 17.093 & 13.379 & 12.044 & 14.728 & 13.672 & 13.056 \\
    	  $r_p/r_e$ & 0.52 & 0.527 & 0.524 & 0.568 & 0.574 & 0.578 & 0.558 & 0.447 & 0.494 & 0.522 & 0.543 \\ \hline
    	\end{tabular}}
    \caption{Astrophysical properties of the neutron, quark and BEC stars rotating at  Keplerian frequencies, for $\rho_c=10^{15}$ g/cm$^3$.}\label{table2}
     \end{table}
     \end{center}
     \end{widetext}

     The  most massive stable star for this model, with mass $M\approx 3.5M_{\odot}$ is obtained for the STOS0 EOS. The BEC50 star has a comparable, but somewhat lower mass, of around $3.2M_{\odot}$. The RMFstiff and the quark equation of state $Q$ have Keplerian masses of the order of $2.8M_{\odot}$, greater than the Keplerian mass of the BEC15 EOS, $M=2.31M_{\odot}$. The smallest Keplerian mass is obtained for the BEC15 EOS. The BEC stars have the biggest equatorial radii, ranging between 15 and 17 km. For the neutron and quark star EOSs the highest equatorial radius is obtained for the STOS0 EOS, but which is still smaller than the equatorial radius of the BEC15 model. The equatorial radius of the RMSsoft EOS almost coincides with the Keplerian equatorial radius of the BEC15 EOS, with the RMSsoft EOS having a much greater mass than the BEC15 EOS one. The Keplerian model of the STOS0 EOS has the greatest moment of inertia, and angular momentum per unit mass, with the BEC15 having the smallest values for these quantities.

     \subsubsection{Models with fixed central density and $r_p/r_e=0.85$}

     The astrophysical parameters of the neutron, quark and BEC stars with a fixed central density $\rho _c=10^{15}$ g/cm$^3$, and a fixed ratio of the polar to the equatorial radius $r_p/r_e$ are presented in Table~\ref{table3}.
     \begin{widetext}
     \begin{center}
      \begin{table}
      \scalebox{1} {
     	\begin{tabular}{| l| c| c| c| c| c| c| c| c| c| c| c| c| }
     	  \hline \hline
     	 EOS & APR & BBBAV14 & BBBParis & BEC15 & BEC30 & BEC50 & DH & Q & RMFsoft & RMFstiff & STOS0 \\ \hline
 	   $\rho_c [10^{15}$ g/cm$^3]$ & 1.00 & 1.00 & 1.00 & 1.00 & 1.00 & 1.00 & 1.00 & 1.00 & 1.00 & 1.00 & 1.00 \\
 	  $M [M_\odot]$ & 1.539 & 1.322 & 1.334 & 1.116 & 2.068 & 2.907 & 1.536 & 1.924 & 1.393 & 2.266 & 2.979 \\
 	  $M_0 [M_\odot]$ & 1.735 & 1.464 & 1.478 & 1.196 & 2.286 & 3.264 & 1.702 & 2.265 & 1.537 & 2.626 & 3.557 \\
 	  $R_e$ [km] & 11.723 & 11.507 & 11.569 & 12.825 & 15.803 & 17.899 & 12.619 & 12.102 & 12.686 & 14.145 & 15.395 \\
 	  $\Omega [10^3$ s$^{-1}]$ & 5.273 & 5.030 & 4.995 & 4.132 & 4.201 & 4.221 & 4.911 & 5.359 & 4.385 & 4.951 & 5.207 \\
 	  $\Omega_p [10^3$ s$^{-1}]$ & 11.247 & 10.761 & 10.725 & 8.417 & 8.263 & 8.034 & 10.036 & 11.974 & 9.576 & 10.195 & 10.153 \\
 	  $T / W [10^{-2}]$ & 4.702 & 4.517 & 4.536 & 3.808 & 3.917 & 3.996 & 4.286 & 5.245 & 4.51 & 4.879 & 5.211 \\
 	  $cJ/GM^2_\odot$ & 1.006 & 0.747 & 0.762 & 0.517 & 1.643 & 3.239 & 0.96 & 1.628 & 0.844 & 2.169 & 3.86 \\
 	  $I [10^{45}$ g cm$^2]$ & 1.677 & 1.305 & 1.341 & 1.100 & 3.436 & 6.745 & 1.718 & 2.670 & 1.693 & 3.850 & 6.518 \\
 	  $\Phi_2 [10^{43}$ g cm$^2]$ & 10.785 & 8.946 & 9.236 & 7.664 & 17.988 & 29.527 & 10.39 & 16.897 & 12.332 & 21.675 & 32.598 \\
 	  $h_+$ [km] & 0.00 & 0.00 & 0.00 & 0.00 & 0.00 & 3.286 & 0.00 & 0.00 & 0.00 & 0.00 & 5.403 \\
 	  $h_-$ [km] & 5.896 & 3.947 & 4.034 & 0.689 & 7.105 & 13.618 & 4.835 & 0.00 & 0.00 & 11.145 & 17.492 \\
 	  $\omega_c/\Omega [10^{-1}]$ & 4.954 & 4.419 & 4.427 & 3.755 & 5.634 & 6.998 & 4.875 & 5.613 & 4.3 & 6.135 & 7.366 \\
 	  $r_e$ [km] & 9.25 & 9.405 & 9.447 & 11.08 & 12.507 & 13.173 & 10.177 & 8.93 & 10.476 & 10.433 & 10.387 \\
 	  $r_p/r_e$ & 0.85 & 0.85 & 0.85 & 0.85 & 0.85 & 0.85 & 0.85 & 0.85 & 0.85 & 0.85 & 0.85 \\ \hline
     	\end{tabular}}
      \caption{Astrophysical parameters for neutron, quark and BEC stars for $\rho_c=10^{15}$ g/cm$^3$ and $r_p/r_e=0.85$.}\label{table3}
      \end{table}
      \end{center}
      \end{widetext}

      Similarly to the previous cases, the higher mass for stars having fixed central density and a polar to equatorial radius ratio $r_p/r_e=0.85$ is obtained for the STOS0 EOS, $M=2.9792M_{\odot}$. The mass of the BEC50 EOS model is lower, with a numerical value of around $2.9M_{\odot}$. The lowest mass value is obtained for the BEC15 EOS, $M=1.11M_{\odot}$. The mass of the BEC30 EOS model, $M=2.068M_{\odot}$ is significantly smaller than the masses obtained for the Q EOS, with $M=1.924M_{\odot}$, and for the RMFstiff EOS, having a corresponding mass of $M=2.266M_{\odot}$. The BEC stars have the larger equatorial radii, of the order of 13-18 km, while the DH EOS has an equatorial radius $r_e\approx 12.6$ km, the largest equatorial radius in this class of stellar models. The BEC15 EOS has the smallest moment of inertia and angular momentum per unit mass, with the highest values of these parameters obtained for STOS0 EOS. However, the angular momentum per unit mass of the BEC50 EOS just slightly exceeds the angular momentum per unit mass of the RMSstiff EOS, but it is smaller than the $cJ/GM$ value of the quark EOS Q.

\section{Electromagnetic and thermodynamic signatures of accretion
disks around Bose-Einstein Condensate stars}\label{sect4}

In the present Section we consider the electromagnetic signatures of the accretion disks around neutron, quark and BEC stars. We consider a comparative study involving three distinct classes of stellar models. The first model corresponds to accretion disks formed around compact general relativistic objects with fixed masses, of the order of $M\approx 1.8M_{\odot}$, rotating at an angular speed of $\Omega \approx 5\times 10^3$ s$^{_1}$. The second case corresponds to accretion disks formed around stars rotating at the maximal Keplerian frequency. And, finally, we also consider a third class of models, in which the accretion disk is located around a star with fixed central density and polar to equatorial radius ratio. For all these three cases we consider the disk emissivity properties, which are strongly dependent of the equation of state of the dense matter inside the star.

\subsection{Electromagnetic spectrum from accretion disks around rotating neutron, quark and BEC stars with fixed mass and angular velocity}

\begin{figure}[tbp]
\centering
\includegraphics[width=8cm]{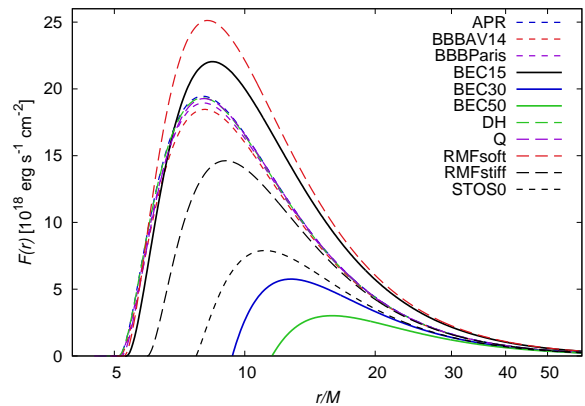}
\caption{Electromagnetic fluxes from accretion disks gravitating around compact general relativistic objects, having different equations of state, with mass $M=1.8M_{\odot}$, and rotating at an angular velocity of $\Omega =5 \times 10^3\; {\rm s}^{-1}$.}
\label{fig4}
\end{figure}

\begin{figure}[tbp]
\centering
\includegraphics[width=8cm]{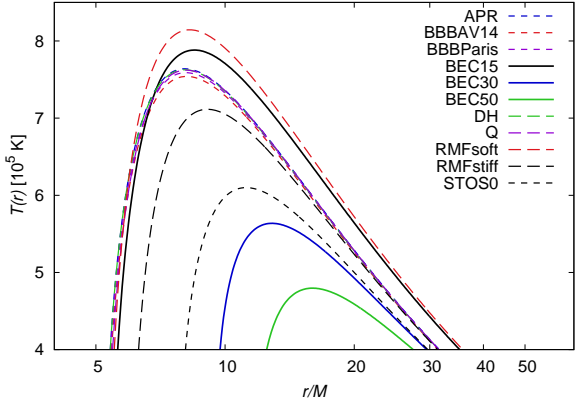}
\caption{Temperature distribution of the accretion disks around  compact general relativistic objects, having different equations of state, with mass $M=1.8M_{\odot}$, and rotating at an angular velocity of $\Omega ==5 \times 10^3\; {\rm s}^{-1}$.}
\label{fig5}
\end{figure}

\begin{figure}[tbp]
\centering
\includegraphics[width=8cm]{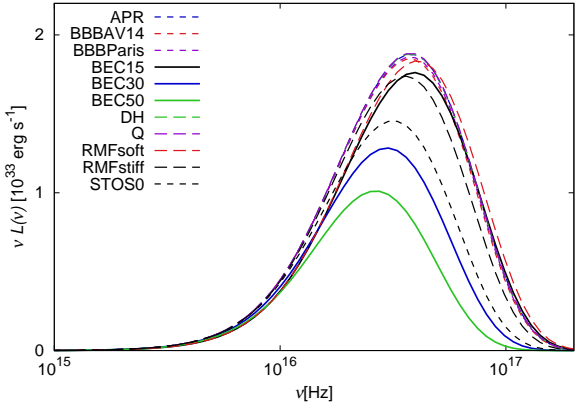}
\caption{Luminosity of the accretion disks around  compact general relativistic objects, having different equations of state, with mass $M=1.8M_{\odot}$, and  rotating at an angular velocity of $\Omega =5 \times 10^3\; {\rm s}^{-1}$.}
\label{fig6}
\end{figure}

We begin our analysis of the electromagnetic signatures of accretion disks around compact general relativistic objects by considering the case of stars with fixed mass, $M\approx 1.8M_{\odot}$, and angular velocity of $\Omega \approx 5 \times 10^3\; {\rm s}^{-1}$.
The variations of the electromagnetic flux, disk temperature and luminosity for neutron, quark and BEC stars with fixed mass and angular velocity are presented in Figs.~\ref{fig4}-\ref{fig6}.

As one can see from the Figures, the BEC stars form a distinct group with respect to the neutron and quark stars included in the study. The flux emitted by the accretion disks, presented in Fig.~\ref{fig4},  is the smallest for the BEC50 and BEC30 equations of state, respectively.  The inner disk edge for the BEC50 and BEC30 EOSs is located at $r/M\simeq 12$ and $r/M\simeq 9$, respectively, at the highest distance from the central object for all considered stars. The maximum value of the flux is obtained for EOS RMFsoft, and this maximum flux value is about twelve and five times bigger than the maximum flux values from the BEC50 and BEC30 EOSs. The flux emitted by the star with EOS BEC15 has high values of the flux, comparable with those from EOS RMFsoft, but this can be explained by the lower mass of the star, $1.541M_{\odot}$, and the corresponding scaling of the flux. However, for the BEC15 EOS the inner edge of the disk is located at a distance of around $5.5\times r/M$ from the central object.

 The temperature distribution in the disk, shown in Fig.~\ref{fig5}, generally follows the same distribution as for the flux profiles, with the BEC stars having some specific distinctive features. The lowest maximum disk temperature is obtained for the BEC50 and BEC30 EOSs, with the maximum temperature located at around $17\times r/M$ and $13\times r/M$, respectively. For most of the neutron and quark stars the maximum disk temperature is reached at $r/M\approx 8$. Some specific distinctive features also appear for EOS STOS0, with a maximum temperature located at around $11\times r/M$, relatively close to the temperature maximum for EOS BEC30. The highest temperature of the disk is reached by EOS RMFsoft.

 The maximum value of the luminosity $\nu L(\nu )$ of the disk, presented in Fig.~\ref{fig6}, is blue shifted for the BEC50 and BEC30 EOSs, and it is reached at a smaller frequency. The maximum values of the luminosity are smaller for these BEC EOSs by a factor of around 1.8. Hence a shift in the position of the luminosity maximum, and a lower value of the luminosity give two specific signatures that could help identify BEC 50 and BEC30 stars via the study of the luminosity of accretion disks around stars with known physical parameters.

\subsection{Electromagnetic signatures of accretion disks gravitating around neutron, quark and BEC stars at Keplerian frequencies}

\begin{figure}[tbp]
\centering
\includegraphics[width=8cm]{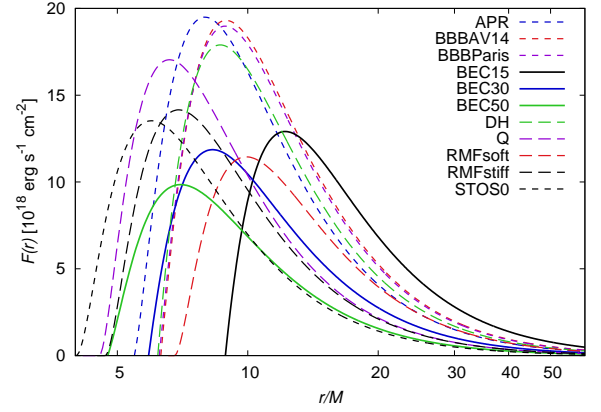}
\caption{Electromagnetic fluxes from accretion disks around compact general relativistic objects with different equations state rotating at Keplerian frequencies.}
\label{fig7}
\end{figure}

\begin{figure}[tbp]
\centering
\includegraphics[width=8cm]{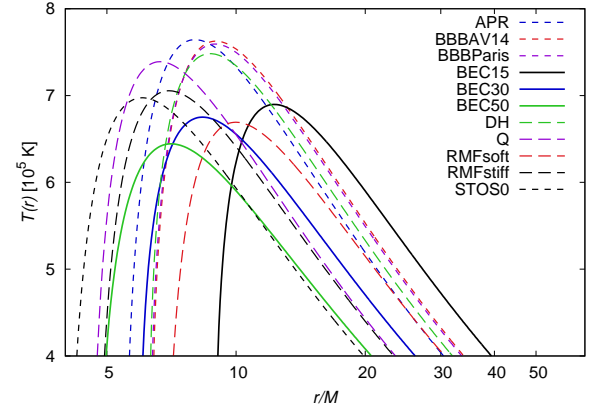}
\caption{Temperature distribution of the accretion disks around  compact general relativistic objects with different equations of state rotating at Keplerian frequencies.}
\label{fig8}
\end{figure}

\begin{figure}[tbp]
\centering
\includegraphics[width=8cm]{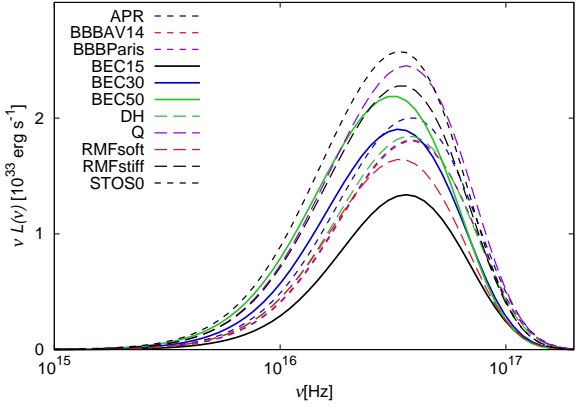}
\caption{Luminosity of the accretion disks around  compact general relativistic objects with different equations of state rotating at Keplerian frequencies. }
\label{fig9}
\end{figure}

As a second example of specific electromagnetic signatures from accretion disks around neutron, quark and BEC stars we consider the case of accretion disks formed around compact stars rotating at Keplerian frequencies. The corresponding electromagnetic fluxes, the disk temperature distribution, and the luminosities are presented in Figs.~\ref{fig7}-\ref{fig9}.

From the point of view of the maximum of the flux emission from accretion disks around neutron, quark and BEC stars rotating at Keplerian frequencies, the stellar models considered in the present study can be roughly divided in two classes: disks with high flux values, and disks with low flux values. As can be seen from Fig.~\ref{fig7}, EOSs APR, BBBAV14, Q and DH have relatively similar maximum flux values, which exceeds the maximum flux values of EOSs RMFstiff, RMFsoft, STOS0, and BEC50, BEC30, and BEC15, respectively. Hence from the point of view of the flux maximum the BEC equations of state present some similarities with the RMFstiff, RMFsoft and STOS0 equations of state. However, important differences do appear in the localization of the position of the maximum of the flux. While for most of the equations of state the maximum is located in the region $7-10\times r/M$, the position of the maximum flux for EOS BEC15 is at around $13\times r/M$. For this equation of state the inner edge of the disk is located at $9\times r/M$, while the inner edges of the BEC50 and BEC30 EOSs are located in positions similar to the other neutron and quark matter equations of state. Hence the determination of the position of the flux maximum from an accretion disk could give a clear indication about the nature of the central compact object. The temperature distribution of the disks around stars rotating at Keplerian frequencies, shown in Fig.~\ref{fig8}, also indicates the existences of two distinct types of stars, with the BEC stars belonging to the lower temperature group. However, the position of the temperature maximum gives a very clear signature of the nature of the equation of state of the compact object. The position of the temperature maximum shifts towards higher distances from the central object with the decrease of the parameter $a/m^3$.

The luminosity of the disk, plotted in Fig.~\ref{fig9}, shows that the maximum of the luminosity is reached at about the same frequency for all considered equations of state. However, the luminosity maximum has different values for different EOSs, with the BEC15 EOS having the smallest value. The BEC50 and BEC30 EOSs have similar luminosity values as for the other considered neutron and quark matter equations of state.

\subsection{Electromagnetic signatures of accretion disks around neutron, quark and BEC stars with fixed central density and $r_p/r_e=0.85$}

\begin{figure}[tbp]
\centering
\includegraphics[width=8cm]{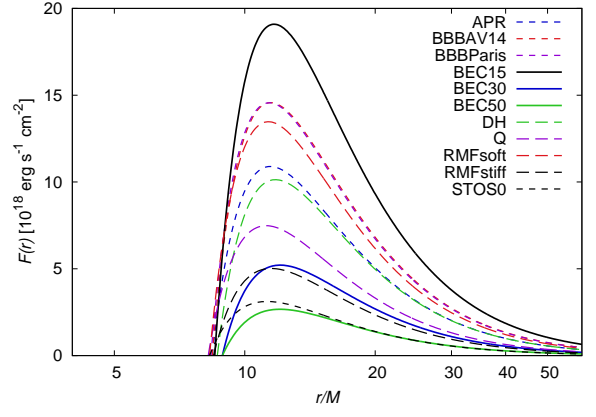}
\caption{Electromagnetic fluxes from accretion disks around compact general relativistic objects with fixed central density $\rho_c=10^{15}$ g/cm$^3$ and $r_p/r_e=0.85$.}
\label{fig10}
\end{figure}

\begin{figure}[tbp]
\centering
\includegraphics[width=8cm]{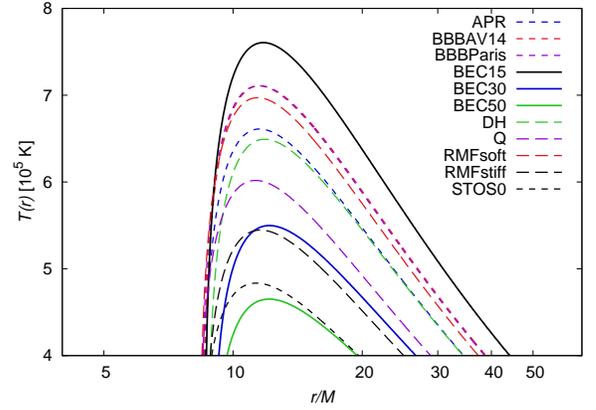}
\caption{Temperature distribution of the accretion disks around  compact general relativistic objects with fixed central density $\rho_c=10^{15}$ g/cm$^3$ and $r_p/r_e=0.85$.}
\label{fig11}
\end{figure}

\begin{figure}[tbp]
\centering
\includegraphics[width=8cm]{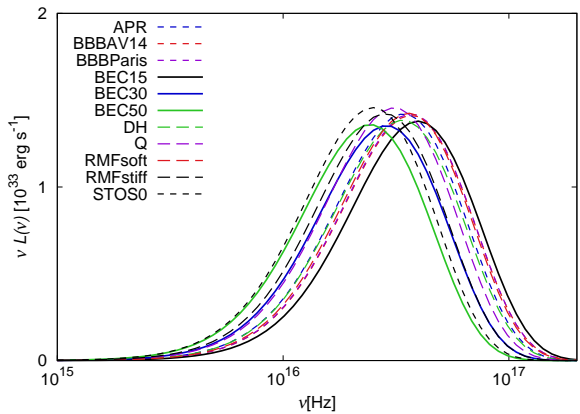}
\caption{Luminosity of the accretion disks around  compact general relativistic objects with fixed central density $\rho_c=10^{15}$ g/cm$^3$ and $r_p/r_e=0.85$.}
\label{fig12}
\end{figure}

Finally, we consider the electromagnetic properties of the accretion disks around neutron, quark and BEC stars with fixed central density and ratio of the polar to the equatorial radius $r_p/r_e=0.85$. All the stellar models have high angular speeds $\Omega $.  The variation with respect to $r/M$ of the emitted fluxes, the temperature distribution of the accretion disks, and their luminosities are presented in Figs.~\ref{fig10}-\ref{fig12}.

From the point of view of the electromagnetic flux distribution, presented in Fig.~\ref{fig10}, the fluxes can be classified in three groups, having high flux values (EOSs Q and APR), medium values (EOSs RMFstiff, RMFsoft, STOS0, DH, BBBAV14, BBBParis), and low flux values, with all the BEC EOSs belonging to this latter group. It is interesting to note that for all BEC EOSs the maximum value of the flux has (approximately) the same value. However, these maxima are located at different $r/M$, with the maximum of the BEC15 EOS positioned at around $15\times r/M$ from the central object. The maximum value of the flux of the BEC stars is around three to four  times lower than from the stars in the first group, with maximum fluxes (EOSs Q and APR), and around two times smaller as compared to the maximum flux values for the EOSs of the second group. The temperature distribution, shown in Fig.~\ref{fig11}, follows the same pattern as the flux, with the maximum of the disk temperature situated at around $18\times r/M$ for EOS BEC15. The maximum temperature value is very similar for all three BEC equations of state. The luminosities of the disks, plotted in Fig.~\ref{fig12}, shows that the maximum of the function $\nu L(\nu)$ is reached at roughly the same frequency for all considered EOSs. However, there is a clear theoretical  difference in the absolute value of the luminosity maximum, with the BEC15 EOS having the smallest luminosity. Even that EOSs BEC 50 and BEC30 have disk luminosities comparable with some neutron star models, they still belong to the class of low luminosity disks.

\subsection{Efficiency of radiation emission from accretion disks around neutron, quark and BEC stars}

 An important observable physical parameter of the disk, which could help to observationally distinguish between different classes of neutron, quark and BEC stars, is the efficiency $\epsilon$ of the conversion of the accreting mass into radiation, given by Eq.~ (\ref{epsilon}). The numerical values of $\epsilon$ show the efficiency of the energy generating mechanism of the mass accretion \cite{Ko}. From a physical point of view, the binding energy $\widetilde{E}_{ms}$,  or $\widetilde{E}_{e}$, represents the amount of energy released by the matter leaving the marginally stable orbit, or the inner edge of the disk, touching the surface of the star, and being transferred to the star. The radii of the inner disk edges for different equation of state and the efficiency of the radiation emission is presented, for the three stellar models considered in the present paper, in Table~\ref{table4}.

 \begin{widetext}
\begin{center}
\begin{table}
   \scalebox{1} {
     	\begin{tabular}{| l| c| c| c| c| c| c| c| c| c| c| c| c| }
     	  \hline \hline
     	  EOS & APR & BBBAV14 & BBBParis & BEC15 & BEC30 & BEC50 & DH & Q & RMFsoft & RMFstiff & STOS0 \\ \hline
     	  Kepler $\;\;\;\;\;\;\;\;\;\;\;\;\;\;\;r_{in}$ [km] & 15.495 & 15.280 & 15.408 & 16.831 & 20.132 & 22.291 & 16.251 & 17.171 & 17.801 & 18.438 & 19.148 \\
     	  $\epsilon$ & 0.0729 & 0.0659 & 0.0662 & 0.0498 & 0.0695 & 0.0796 & 0.0675 & 0.0807 & 0.0596 & 0.0817 & 0.0905 \\ \hline
     	  $\frac{r_p}{r_e}=0.85$ $\;\;\;\;\;\;\;\;\;\;r_{in}$ [km] & 19.067 & 16.330 & 16.445 & 14.030 & 26.987 & 37.927 & 19.482 & 23.327 & 17.028 & 28.148 & 36.294 \\
     	  $\epsilon$ & 0.0527 & 0.0524 & 0.0527 & 0.0512 & 0.0503 & 0.0505 & 0.0514 & 0.0539 & 0.0528 & 0.0527 & 0.0538 \\ \hline
     	  $\Omega =5\times 10^3\;{\rm s}^{-1}$ $r_{in}$ [km] & 12.171 & 13.003 & 12.715 & 11.952 & 24.917 & 30.661 & 12.204 & 11.979 & 11.344 & 15.812 & 21.038 \\
     	  $\epsilon$ & 0.0663 & 0.0670 & 0.0671 & 0.0646 & 0.0479 & 0.0389 & 0.0658 & 0.0635 & 0.0668 & 0.0635 & 0.0537 \\ \hline
     	\end{tabular}}
      \caption{The radius of the inner disk edge $r_{in}$, and the efficiency $\epsilon$ of the matter-radiation conversion for stars in Keplerian rotation, $r_p/r_e=0.85$, and $\Omega =5\times 10^3$ s$^{-1}$.}\label{table4}
      \end{table}
      \end{center}
      \end{widetext}

   The first two lines of Table~\ref{table4} contain the conversion efficiency of the compact neutron, quark and BEC stars rotating at Keplerian frequencies. In this case the values of $\epsilon $ for the BEC stars are in the range of 5\% to 8\%, having values comparable to the $\epsilon $ values for neutron and quark stars. The highest efficiency, 9\%, is obtained for the STOS0 EOS, while the BEC50 EOS has a conversion efficiency close to that of the quark stars. The smallest $\epsilon $ value is obtained for the BEC15 EOS, showing that these stars are less efficient engines for the conversion of the accreted mass into outgoing radiation.

The values of $\epsilon $ for the compact general relativistic objects with fixed central density and $r_p/r_e$ are given in the second two lines of Table~\ref{table4}. The values of $\epsilon $ are slightly lower, as compared to the Keplerian rotation case, with the highest value for $\epsilon $ obtained for the STOS0 EOS. The smallest value of $\epsilon $ is found for the BEC15 EOS. However, in this case the efficiency of the BEC50 EOS is significantly smaller than the efficiency of the Q EOS, being slightly higher than $\epsilon $ for APR EOS.

The last two lines in Table~\ref{table4} show the numerical values of $\epsilon$ for disks around compact objects with mass $M=1.8M_{\odot}$, rotating with an angular velocity of $\Omega \approx 5\times 10^3$ s$^{-1}$. While the efficiency of the radiation conversion is almost the same for EOSs APR, BBBAV14, BBBParis, DH, Q, RMFsoft, and RMFstiff, it has some lower values for the BEC15 EOS, and much lower values for EOSs BEC30 and BEC50. Relatively low values have been also obtained for the STOS0 EOS. The efficiencies of 3.89\% obtained for the BEC15 and BEC50 stars are the lowest accretion disk efficiencies obtained in the present study. The inner edges of the accretion disks around BEC stars are located at a much bigger distance from the
central object as compared to the other classes of neutron and quark stars, reaching a value of 31 km for the BEC50 EOS. Such a far away located inner edge does explain the low efficiency of the corresponding accretion disk.

\section{Discussions and final remarks}\label{sect5}

The possible existence of some forms of Bose-Einstein Condensates in compact general relativistic objects, or the existence of pure Bose-Einstein Condensate stars, represents an intriguing, and interesting, possibility, for which a lot of theoretical evidence has been provided. From an observational point of view the most important differences between standard neutron or quark stars are represented by the differences in mass and radius. If the masses and radii of the compact general relativistic object could be measured with high accuracy, these measurements would put very strong {\it direct} constraints on the equation of state of the dense star. However, presently, there are very few precise determinations of both the mass and radius of a compact object. Therefore, in the present paper we have proposed, and preliminary investigated, an alternative {\it indirect} method, that could help observationally distinguishing between different classes of compact objects, and their equations of state. This method is based on the information extracted from observations of the basic physical properties of
matter forming thin accretion disks around rapidly rotating neutron, quark and Bose-Einstein Condensate stars.

Due to the presence of a strong gravitational field, all the astrophysical quantities related to the observable properties of the accretion disks, are dependent, and can be obtained from the metric of the central compact object \cite{Ko}. Due to the major differences in the exterior space-time
geometry, neutron, quark and Bose-Einstein condensate stars  show, at least on the theoretical level,  some very important distinct signatures with respect to the disk properties.  Therefore, the observational procedure of the analysis of the electromagnetic radiation of accretion disks may allow to discriminate between neutron, quark and Bose-Einstein Condensate stars, by giving some specific distinct signatures that could  differentiate between compact objects described by different equations of state. In the present paper we have obtained the physical parameters of the disk - effective potential, flux and emission spectrum profiles - for several equations of state of the neutron, quark and BEC matter, respectively.

As one can see from the flux integral in Eq.~(\ref{F}), as well as from the explicit expressions of the specific energy, specific angular momentum and angular velocity, given by Eqs.~(\ref{rotE}), (\ref{rotL}) and (\ref{rotOmega}), respectively,  the rather different characteristics of the radial flux distribution over the accretion disk, the disk spectra and the conversion efficiency are due to the important differences between the metric potentials of the neutron, quark and Bose-Einstein Condensate stars, respectively. Even if the total mass and the angular velocity are the same for each type of the rotating central object (neutron, quark or Bose-Einstein Condensate star), with the stars having similar values of $\Omega$, $\widetilde E$ and $\widetilde L$, the radiation properties of the accretion disks around these compact general relativistic objects exhibit observable differences \cite{Ko}. The physical reason for these differences is that the proper volume, and in turn the function $\sqrt{-g}$, used in the calculation of the flux integral, is strongly dependent on the behavior of the metric component $g_{rr}=(\partial \overline{r}/\partial r)^2 g_{\overline{r}\overline{r}}$, and therefore on the geometry of the space-time. The latter expression contains the derivatives with respect to the radial coordinate $r$ of the metric components $\rho(r)$ and $\gamma(r)$, respectively, via the coordinate transformation between the coordinates $\overline{r}$ and $r$, which are extremely sensitive to the slope of the functions $\rho(r)$ and $\gamma(r)$ \cite{Ko}. Therefore, although the inner edges of the disks are located at almost the same radii, the maximum amplitudes and the numerical values of the energy fluxes emerging form the disk surface, and propagating in any solid angle, may show considerable differences for different equations of state of the neutron, quark and Bose-Einstein Condensate matter. These essentially geometrical effects  also give the distinctive features in the disk spectra for the various types of central stars.

In our preliminary and idealized theoretical study of the accretion disk properties around Bose-Einstein Condensate stars we have found a number of observational signatures distinguishing this class of stars from the neutron and quark stars. These specific properties are the distinct positions of the maxima of the flux, of the temperature distribution, and of the luminosity of the disk, the position of the inner edge of the disk, and the radiation efficiency conversion. In all three different classes of rotating stars we have analyzed these signatures do appear distinctly. Moreover, Bose-Einstein Condensate stars have a mass-radius relation that can also help in discriminating them with respect to other classes of neutron and quark stars.

In the present paper, which represents a first step in the investigation of the complex astrophysical problem of the radiation emission from accretion disks around compact objects with different nuclear equations of state, we made the fundamental assumption that the inner edge of the disk is located at the ISCO radius. This means that the ISCO position is only determined by the space-time geometry around the compact object. However, a number of important physical factors can affect the position and location of the inner edge of the disk. For example, in the case of neutron stars,  the inner edge of the disk is usually set by the magnetosphere around the star,  and it is not determined by the metric of the space-time only. But, in order to simplify the discussion of the complex physics of the accretion  disks, and by taking into account the preliminary and purely theoretical nature of this study, in the present paper we consider only an ideal case, which allows us to start our analysis from very simple physical and astrophysical considerations. As already mentioned, an important physical parameter that strongly influences the ISCO position is the magnetic field of the compact central object, and its magnetosphere.  The magnetic field lines originating from the central object
can have a considerable effect on the evolution of the on the accretion processes, and on the disk structure. The effects of magnetic fields on the accretion processes in a compact, spherically symmetric geometry of Schwarzschild type (which describes the exterior geometry for both black holes and compact stars) were considered in \cite{Kov}. The magnetic field was assumed to be asymptotically uniform, and axisymmetric tidal
structures were also taken into account. Due to the presence of the magnetic field and of the tidal perturbations, the accretion disk shrinks in size, and the marginally stable orbits shift towards the central object. On the other hand the presence of the magnetic perturbation leads to an increase in the disk radiation intensity from the accretion disk. However, the position of the maximum of the radiation
does not change, and the radiation spectrum is blue-shifted. Interestingly, the conversion efficiency decreases due to the presence of the magnetic fields and of the tidal perturbations. Alternative studies of the effects of the neutron star's magnetic field on the exterior metric and the position of ISCO's can be found in \cite{stut}.

In  the present paper we have proposed a method for discriminating between different types of compact objects that was proposed, and observationally tested, in the case of the black holes. Black holes have an event horizon (or a very special surface with similar properties) that prevent them to emit any form of electromagnetic radiation \cite{Nar}. This makes the radiation emission from the disk, and due to accretion processes, to be the dominant electromagnetic energy emission mechanism, thus allowing the possibility of determining the central black hole properties from the radiation spectrum of the disk, without any possible interference of the electromagnetic signals from the black hole. On the other hand, neutron stars have a solid radiation emitting surface (crust), which makes very difficult to distinguish between the radiation of the star and of the disk itself. That's why the continuum-fitting method was not applied for the study of neutron stars \cite{Nar}, and up to now there is no published observational study of the possibility of testing the neutron star equation of state by using the electromagnetic spectrum of a thin disk. On the other hand we would like to point out that the electromagnetic emissivity of the zero temperature gravitationally bounded Bose-Einstein Condensate matter, in which all particles are in the quantum ground state, described by a single wave function, is very low \cite{em}. This is because the effective plasma frequency of the condensed mater is very high. Therefore we may assume that the radiation emissivity of the BEC stars is very low, and they are very "black". Consequently,  the electromagnetic emissivity of the  BEC star - disk system may be dominated by the disk emission. This situation is similar to the case of quark stars. Since quark matter has a very high plasma frequency $\omega _p$, the  photon emissivity of strange quark stars is very low  \cite{qu}. This is due to the fact that the
propagation of electromagnetic waves having frequencies lower than $\omega _p$  is exponentially damped.
Hence,  only photons produced just below a few fermi from the surface, with outwards pointing momenta
can be emitted by the strange star. Hence the equilibrium photon emissivity from a strange star is negligible small, as compared to the black body one. Moreover, the spectrum of the emitted equilibrium photons is very hard, with $\hbar \omega > 20$ MeV \cite{qu}.
Of course BEC stars, as well as the quark stars,  may have a crust (solid surface), representing a powerful source of electromagnetic radiation. The presence of such a crust would further complicate the possibility of discriminating between BEC stars and the other types of compact general relativistic objects.

Another important point we would like to stress is that from an observational point of view  not all the parameters of the BEC star-disk system can be determined independently from astrophysical observations. This means that a degeneracy between the equation of state and the nature of the compact object, and the values of some physical parameters of the model, could always exist. A full test of the BEC equation of state would require the precise knowledge of the mass, radius, and spin of the central object, as well as all the electromagnetic disk properties. In this ideal case, extremely difficult to be achieved from observational point of view, the fitting of the observational data could provide a convincing test of the nature of the central object. Probably in the near future no such increase in the precision of the astrophysical observations will be achieved, thus making the direct determination of the EOS of the nuclear or condensed matter to be beyond the present observational capabilities. On the other hand one could expect an increase in the determination of the masses and radii for various compact objects, and these measurements may give some hints on the true nature of the equation of state of the dense matter.

In conclusion, once the precision of the astrophysical data is drastically increased, the observational study of the thin accretion disks around rapidly rotating compact objects, and of their electromagnetic properties (flux, temperature distribution and luminosity),  may provide a powerful tool in distinguishing between different classes of dense stellar objects, as well as for discriminating between the different equations of state of the dense matter.

\section*{Acknowledgments}

      We would like to thank to the anonymous referee for comments and suggestions that helped us to significantly improve our manuscript. The authors are very grateful to Dr. Gabriela Mocanu for her constant support, help, and useful comments. BD acknowledges the support of "Babes-Bolyai" University Cluj-Napoca through a Research Excellence scholarship. This work is partially supported by a grant of the Romanian National Authority of Scientific Research, Program for research - Space Technology and Advanced Research - STAR, project number 72/29.11.2013.

\end{document}